\documentclass[12pt]{article}
\newtheorem{definition}{Definition}
\newtheorem{proposition}{Proposition}
\newtheorem{lemma}{Lemma}

\newtheorem{theorem}{Theorem}
\usepackage{ytableau}
\usepackage[all]{xy}
\usepackage{tikz, tikz-cd}
\usepackage{mathdots}
\usepackage{savesym}
\usepackage{cite}
\usepackage{wasysym}
\usepackage{amscd}
\usepackage{graphicx}
\usepackage{pifont}
\usepackage{float}
\usepackage{subfig}
\usepackage{multicol}
\usepackage{amsfonts}
\usepackage{picture}
\usepackage{amssymb}
\savesymbol{iint}
\savesymbol{iiint}
\usepackage{amsmath}
\restoresymbol{TXF}{iint}
\restoresymbol{TXF}{iiint}
\usepackage{color}
\usepackage{clay}
\usepackage{hyperref}
\usepackage{slashed}
\hypersetup{colorlinks=true}
\hypersetup{linkcolor=black}
\hypersetup{citecolor=black}
\hypersetup{urlcolor=black}
\usepackage{setspace}
\usepackage{multirow}
\usepackage{wrapfig}
\usepackage{verbatim}
\numberwithin{equation}{section}

\usepackage{suffix}

\begin{document}

\institution{Fellows}{\centerline{${}^{1}$Society of Fellows, Harvard University, Cambridge, MA, USA}}
\institution{HarvardU}{\centerline{${}^{2}$Jefferson Physical Laboratory, Harvard University, Cambridge, MA, USA}}

\title{Counting Trees in Supersymmetric Quantum Mechanics}

\authors{Clay C\'{o}rdova\worksat{\Fellows}\footnote{e-mail: {\tt cordova@physics.harvard.edu}}  and Shu-Heng Shao\worksat{\HarvardU}\footnote{e-mail: {\tt shshao@physics.harvard.edu}} }

\abstract{We study the supersymmetric ground states of the Kronecker model of quiver quantum mechanics.   This is the simplest quiver with two gauge groups and bifundamental matter fields, and appears universally in four-dimensional $\mathcal{N}=2$ systems.      The ground state degeneracy may be written as a multi-dimensional contour integral, and the enumeration of poles can be simply phrased as counting bipartite trees.  We solve this combinatorics problem, thereby obtaining exact formulas for the degeneracies of an infinite class of models.  We also develop an algorithm to compute the angular momentum of the ground states, and present explicit expressions for the refined indices of theories where one rank is small.}

\date{February 2015}
\maketitle

\begin{spacing}{.999}
\tableofcontents
\end{spacing}

\section{Introduction}
 
One of the most basic problems in any quantum system is to determine the spectrum of stable states.  However, outside the realm of perturbation theory and exactly solvable models the answer to this question is elusive.   One class of examples where progress is possible are those enjoying supersymmetry.  In general, these models are not exactly solvable, yet nevertheless there are states in the spectrum whose existence and properties can be reliably determined.  These supersymmetric models yield a window into non-perturbative physics where strong coupling phenomena may be confronted analytically.  In particular, field theories and gravities with extended supersymmetry provide a class of models where exact spectroscopy results are feasible. 

Motivated by these general considerations, in this work we analyze in detail a non-trivial model of supersymmetric particle spectroscopy.  We consider a non-relativistic quantum mechanical system with two distinct species of (super)particles, and four supercharges.  Each particle carries minimal angular momentum, and electromagnetic charges $\gamma_{1}$ and $\gamma_{2}.$  The low-energy interactions of the system are invariantly characterized by the integral Dirac pairing of the electromagnetic charges
\begin{equation}
\langle \gamma_{1}, \gamma_{2} \rangle =k>0~.
\end{equation}
Our aim is to determine the non-relativistic bound state spectrum formed by $M$ particles of type one, and $N$ particles of type two.  The system and its interactions are encoded in a quiver diagram, known as the \emph{Kronecker} quiver shown below.\footnote{The explicit expression for the Hamiltonian of this system may be found, for instance, in \cite{Denef:2002ru}.}
\begin{figure}[here!]
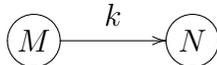

  \centering
\subfloat{
\xy  0;<1pt,0pt>:<0pt,-1pt>::
(-300,0) *+{M}*\cir<10pt>{} ="1",
(-240,0) *+{N}*\cir<10pt>{} ="2",
(-270, -10) *+{k} ="b",
\ar @{->} "1"; "2"
\endxy}
  \caption{The Kronecker quiver with dimension vector $(M,N)$ and $k$ arrows.  The quiver encodes the interaction of $M$ particles of charge $\gamma_{1}$ and $N$ particles of charge $\gamma_{2}$ with $\langle \gamma_{1}, \gamma_{2}\rangle =k.$}\label{fig:kronecker}
\end{figure}
We study only those bound states which preserve all four supercharges.  These are the supersymmetric ground states of this quiver quantum mechanics.  We denote their degeneracy by $\Omega(M,N,k).$   The Kronecker quiver model and the degeneracies $\Omega(M,N,k),$ have been previously studied from a variety of perspectives, including quantum groups \cite{2003InMat.152..349R},  wall-crossing formulas \cite{Kontsevich:2008fj, Gross, Reineke}, spectral networks \cite{Galakhov:2013oja}, and equivariant cohomology \cite{Weist:2009, Weist:2012}.

Our main result, described in detail in \S \ref{kronsec} and \S \ref{sec:check}, is a formula for $\Omega(M,N,k)$ in the special case where the parameter $N$ may be expressed as $N=Mr+1$ for non-negative integer $r$.  It is simplest to state our results in terms of a generating function.  Introduce $F(k,r,x)$ which depends on a formal variable $x$ as
\begin{equation}
F(k,r,x) =(k-r) \sum_{\ell=1}^\infty  {(-1)^{\ell-1}\over \ell} {k\ell \choose r\ell }  x^\ell~.
\end{equation}
and let $[x^j]\{ q(x) \}$ denote the coefficient of $x^{j}$ in a power series  $q(x)$. Then, we find
\begin{equation}
\Omega(M,Mr+1,k)= {1\over (Mr+1)^2}[x^M] \left\{\exp\Big[\, (Mr+1) F(k,r,x)\, \Big] \right\}  ~. \label{resintro}
\end{equation}
In the further limit $r\rightarrow 1$, the generating function simplifies dramatically and we are able to provide a closed form expression
\begin{equation}
\Omega(M,M+1,k)= {k\over (M+1) \left[(k-1)M +k\right] }{ (k-1)^2 M +k(k-1) \choose M}~,
\end{equation}
thus reproducing the results of \cite{Weist:2009}.  In \S \ref{sec:spin}, we also develop an algorithm to compute the angular momentum of the ground states, and apply our algorithm to ground states with small $M$ and arbitrary $N$ in equation \eqref{spinanswers}.

The method that we use to derive \eqref{resintro} is supersymmetric localization \cite{Hwang:2014uwa, Cordova:2014oxa, Hori:2014tda}.  This technique expresses the index $\Omega(M,N,k)$ as a multidimensional contour integral \cite{MR1318878}.  Our key technical results, proved in \S \ref{sec:derivation}, are a systematic combinatorial interpretation of the residues of this integral.  Specifically, we demonstrate that enumeration of poles in the contour integral is equivalent to counting certain bipartite trees.  The computation of the number of such trees is an elementary problem in graph theory which we solve in \S \ref{sec:graph}.  Its solution leads to the index formula \eqref{resintro}.  

\begin{figure}[here!]
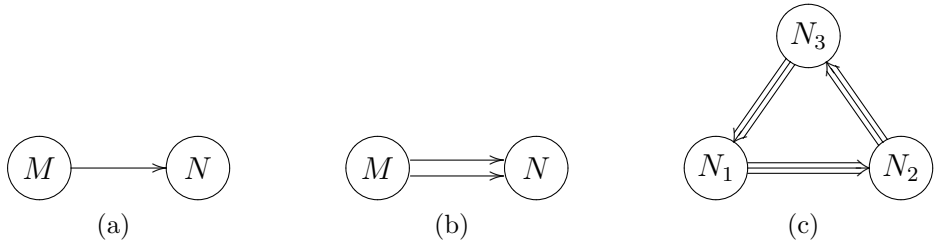

  \centering
  \subfloat[]{
\xy  0;<1pt,0pt>:<0pt,-1pt>::
(-240,0) *+{M}*\cir<12pt>{} ="2",
(-180,0) *+{N}*\cir<12pt>{} ="3",
\ar @{->} "2"; "3",
\endxy}
\hspace{.5in}
  \subfloat[]{
\xy  0;<1pt,0pt>:<0pt,-1pt>::
(-240,0) *+{M}*\cir<12pt>{} ="2",
(-180,0) *+{N}*\cir<12pt>{} ="3",
(-235,3) *+{\phantom{A}}="21",
(-235,-3) *+{\phantom{A}}="22",
(-185,3) *+{\phantom{A}}="31",
(-185,-3) *+{\phantom{A}}="32",
\ar @{->} "21"; "31",
\ar @{->} "22"; "32",
\endxy}
\hspace{.5in}
\subfloat[]{
\xy  0;<1pt,0pt>:<0pt,-1pt>::
(-300,0) *+{N_1}*\cir<12pt>{} ="1",
(-230,0) *+{N_2}*\cir<12pt>{} ="2",
(-265,-50) *+{N_3}*\cir<12pt>{} ="3",
\ar @3{->} "1"; "2"
\ar @3{->}  "3"; "1"
\ar @3{->} "2"; "3"
\endxy}
  \caption{Examples of Kronecker quivers occur in many well-known models.  The case $k=1,$ illustrated in (a), describes the spectrum of the Argyres-Douglas conformal field theory \cite{Argyres:1995jj, Gaiotto:2009hg, Alim:2011ae}.  The case $k=2,$ illustrated in (b), describes the spectrum of $su(2)$ Seiberg-Witten theory \cite{Seiberg:1994rs, Fiol:2000pd}.  The case $k>2$ occurs frequently as a subset of the bound states in many $\mathcal{N}=2$ systems.  For instance, the $k=3$ case in (c) appears in the quiver that describes the bound states of D-branes in type IIA  string theory on local $\mathbb{P}^{2}$ \cite{Douglas:2000qw}.   More generally, all Kronecker quivers with $k>2$ occur as subsectors of the quiver describing bound states in $su(N)$ super-Yang-Mills \cite{Galakhov:2013oja}. }
\label{kronexamps}
\end{figure}

Our results have broad applications in $\mathcal{N}=2$ field theories and supergravity.  Indeed, in a wide class of such systems, the supersymmetric particles and black holes may be captured by a non-relativistic quiver quantum mechanics \cite{ Douglas:2000ah, Douglas:2000qw, Fiol:2000wx, Fiol:2000pd, Denef:2002ru, Denef:2007vg, Alim:2011kw, Manschot:2012rx, Cecotti:2012sf, Chuang:2013wt, Cordova:2013bza}.  The Kronecker model that we study, describes a subset of all such quiver quantum mechanics systems: it encodes the bound states whose electromagnetic charges lie in a sublattice spanned by two primitive charges (see, for example, Figure \ref{kronexamps}).  Moreover, the spectrum of this model illustrates universal physical features such as Regge trajectories \cite{Cordova:2015vma}, and an exponential degeneracy of states \cite{Weist:2009, Galakhov:2013oja, CS2, Tom}.  Finally, the degeneracies $\Omega(M,N,k)$ are also interesting due to the distinguished role that they play in wall-crossing formulas \cite{Kontsevich:2008fj}.

The index formula that we obtain is also of interest in a purely mathematical context.  As we review in \S \ref{kronsec}, the index $\Omega(M,N,k)$ can be interpreted as the Euler characteristic of the moduli space $\mathcal{M}^k_{M,N}$ of stable representations of the Kronecker quiver.  These Kronecker moduli spaces are a natural generalization of Grassmannians.  Our combinatorial interpretation of the cohomology of $\mathcal{M}^k_{M,N}$ is thus reminiscent of classical results in Schubert calculus.

There are a number of significant questions left unanswered by our analysis.  Most glaringly, it would be interesting to extend our result \eqref{resintro} to the general value of parameters $(M,N,k),$ and thereby provide a complete solution to the Kronecker model.  More conceptually, the exponential resummation appearing in \eqref{resintro} is qualitatively similar to the general relationship between Donaldson-Thomas invariants and Gromov-Witten invariants \cite{Gopakumar:1998ii, Gopakumar:1998jq}.  This suggests, perhaps, that the function $F(k,r,x)$ itself admits a direct enumerative meaning.  Finally, it would be satisfying to explain the physical significance of the bipartite trees which play a crucial technical role in this work, perhaps by relating them to attractor flow trees \cite{Denef:2000nb, Denef:2000ar}.  We leave these problems as potential avenues for future investigation.

\section{Kronecker Quivers}
\label{kronsec}
In this section, we give an overview of our main result for the degeneracies $\Omega(M,N,k).$  Complete proofs of all ingredients presented may be found in \S \ref{sec:derivation} and \S \ref{sec:graph}.

We begin in \S \ref{sec:geom} with a review of the geometry underlying the indices of the Kronecker quiver, and explain how the degeneracies $\Omega(M,N,k)$ may be viewed as the Euler characteristic of the Kroncker moduli space.  Next in \S \ref{sec:degen} and \S \ref{subsec:graph} we overview the main steps in our calculation.  In particular, we describe how counting bipartite trees is related to computing indices and in  \S\ref{subsec:tree} we state a theorem which enables us to enumerate all such trees.  Finally, in \S\ref{subsec:result} we assemble the pieces into a formula for  $\Omega(M,N,k),$ in the special case $N=Mr+1.$

\subsection{Kronecker Moduli and the Index}
\label{sec:geom}

In any model of quiver quantum mechanics with four supercharges, the problem of determining the ground state spectrum can be phrased in a completely geometric language in terms of quiver moduli spaces.  In this section we review this connection in the context of the Kronecker quiver.

Consider the Kronecker quiver illustrated in Figure \ref{fig:kronecker}.  The Lagrangian for this system is a gauged $\mathcal{N}=4$ quantum mechanics.  Each node supports a unitary gauge group of ranks $M$ and $N$ respectively with associated vector multiplets, while the arrows of the quiver are bifundametal chiral multiplet matter fields. The explicit expression for the Hamiltonian of this system may be found, for instance, in \cite{Denef:2002ru}.

The system has a classical Higgs branch moduli space $\mathcal{M}^k_{M,N}$ described in a standard way.  The  chiral multiplet fields $\Phi_{i}$ ($i=1,\cdots, k$) have constant expectation values.  Thus, they may be viewed as specifying linear maps 
\begin{equation}
\Phi_{i}: \mathbb{C}^{M}\rightarrow \mathbb{C}^{N}~.
\end{equation}
On the set of possible field expectation values we impose the D-term equations 
\begin{equation}
\sum_{i=1}^{k}\Phi_{i}^{\dagger}\circ\Phi_{i}=\zeta I_{M}~,\hspace{.5in}\sum_{i=1}^{k}\Phi_{i}\circ\Phi_{i}^{\dagger}=\frac{M\zeta}{N} I_{N}~, \label{dterm}
\end{equation}
where in the above, $\zeta>0$ is the Fayet-Iliopoulos parameter,\footnote{When $\zeta<0$ all moduli spaces are empty, illustrating the wall-crossing phenomenon.} and $I_{L}$ is the $L\times L$ identity matrix.  Finally, the locus of solutions to the D-term equation \eqref{dterm} is invariant under the action of the gauge group $U(M)\times U(N)$ acting on the $\Phi_{i}$ via the bifundamental representation.  The desired moduli space $\mathcal{M}^k_{M,N}$ is the quotient space
\begin{equation}
\mathcal{M}^k_{M,N} \equiv \left\{\Phi_{i}~\Bigg\vert \sum_{i=1}^{k}\Phi_{i}^{\dagger}\circ\Phi_{i}=\zeta I_{M}~,\hspace{.25in}\sum_{i=1}^{k}\Phi_{i}\circ\Phi_{i}^{\dagger}=\frac{M\zeta}{N} I_{N}\right\}/U(M)\times U(N)~.
\end{equation}

The Kronecker moduli spaces $\mathcal{M}^k_{M,N}$ have several features which follow directly from their construction as quotients, as well as through the application of Seiberg dualities (quiver mutations \cite{MR0332887, MR0393065}).  We state these facts here.  Throughout this discussion, we assume that the pair $(M,N)$ are coprime, to avoid various subtleties.  
\begin{itemize}
\item The moduli space  $\mathcal{M}^k_{M,N}$ is a smooth compact K\"{a}her manifold (if it is non-empty), with metric inherited via its construction as a K\"{a}her quotient.
\item The complex dimension of $\mathcal{M}^k_{M,N}$ is 
\begin{equation}
\mathrm{dim}(\mathcal{M}^k_{M,N}) = k NM-M^{2}-N^{2}+1~. \label{dimformula}
\end{equation}
In particular, the moduli space is non-empty if and only if the above dimension formula is non-negative.
\item In the Hodge decomposition of the cohomology of $\mathcal{M}^k_{M,N},$ we have \cite{2003InMat.152..349R}
\begin{equation}
p\neq q \Longrightarrow h^{p,q}\left(\mathcal{M}^k_{M,N}\right)=0~. \label{vanishing}
\end{equation}
\item We have the following isomorphisms
\begin{align}\label{mutation}
\begin{split}
&\text{Reflection:}~~~~\mathcal{M}^k_{M,N} \cong \mathcal{M}^k_{N,M}~, \\
&\text{Mutation:}~~~~~\mathcal{M}^k_{M,N}\cong  \mathcal{M}^k_{N,Nk-M}~.
\end{split}
\end{align}
\end{itemize}

According to familiar results in supersymmetric quantum mechanics, the supersymmetric ground states may be extracted from the cohomology of the moduli space.  Moreover, the data of the spin of ground states (an $su(2)$ representation) is determined by the Lefschetz $su(2)$ action on the cohomology. 

Complete information about the ground state spectrum is conveniently packaged into a refined index depending on a fugacity $y$ which encodes the angular momentum
\begin{equation}
\Omega(M,N,k,y) \equiv \sum_{p=0}^d y^{2p-d}\, h^{p,p}(\mathcal{M}^k_{M,N})~, \label{spin}
\end{equation}
where in the above, $d$ denotes the complex dimension of the moduli space given in \eqref{dimformula}.  Up to the overall factor of $y^{-d},$ this index coincides with the Hirzebruch $\chi_{y}$ genus.  In particular, in the specialization $y\rightarrow 1$ the above reduces to the Euler characteristic
\begin{equation}
\Omega(M,N,k) \equiv \Omega(M,N,k,1) = \sum_{p=0}^d  h^{p,p}(\mathcal{M}^k_{M,N}) =\chi\left(\mathcal{M}^k_{M,N}\right)~. \label{nospin}
\end{equation}
One notable feature of both \eqref{spin} and \eqref{nospin} is that the Betti numbers are weighted without signs.  Thus, $\Omega(M,N,k,y),$ which a priori is an index and counts states up to signs, in fact computes the exact degeneracy due to the vanishing result \eqref{vanishing} on the cohomology.\footnote{This is a special case of the ``No-Exotics" conjecture \cite{Gaiotto:2010be, DelZotto:2014bga}.}

The indices \eqref{nospin} and \eqref{spin} are the quantities that we compute in this work.  Our main results concern the Euler characteristic \eqref{nospin}.  In \S \ref{sec:spin} we present an algorithm to determine the complete cohomology generating function \eqref{spin}.

\subsubsection{The Grassmannian as a Kronecker Moduli Space}
\label{sec:grass}

In this section we provide some intuition for Kronecker moduli spaces, by demonstrating that in the special case $(M,N)=(M,1),$ the moduli space reduces to a Grassmannian.  

To illustrate this isomorphism, we first write the maps $\Phi_{i}$ as a row vector
\begin{equation}
\Phi_{i}=\left(\begin{array}{cccc}\phi_{i}^{1} & \phi_{i}^{2} & \cdots & \phi_{i}^{M}\end{array}\right)~.
\end{equation}
The $k$ distinct row vectors may then be assembled into a $k\times M$ matrix $X$ as 
\begin{equation}
X=\left(\begin{array}{cccc}\phi_{1}^{1} & \phi_{1}^{2} & \cdots & \phi_{1}^{M}   \\
\phi_{2}^{1} & \phi_{2}^{2} & \cdots & \phi_{2}^{M} \\
\vdots & \vdots & \ddots & \vdots \\
\phi_{k}^{1} & \phi_{k}^{2} & \cdots & \phi_{k}^{M} \end{array}\right)~.
\end{equation}
The D-term equation \eqref{dterm} asserts that the $M$ columns of this matrix are pairwise orthogonal and that they each have norm $\zeta$.  From this we conclude that matrix $X$ must have rank $M$ and hence in particular the moduli space is empty if $M>k$.

On the other hand if $M\leq k,$ then the columns of the matrix $X$ comprise a unitary frame for an $M$-plane in $\mathbb{C}^{k}.$  The gauge group $U(M)$ acts on $X$ as $X\rightarrow X G$ and hence may be viewed as changing the unitary frame for the $M$-plane.  We conclude that the moduli space is a Grassmannian 
\begin{equation}
\mathcal{M}^{k}_{M,1} \cong Gr(M,k)~.
\end{equation}

The cohomology of the Grassmannian is well-known, and via \eqref{spin} we are able to write the generating function
\begin{equation}
\Omega(M,1,k,y)=\begin{cases}\frac{y^{M(M-k)}\prod_{i=1}^{k}(1-y^{2i}) }{\prod_{i=1}^{M}(1-y^{2i})\prod_{i=1}^{k-M}(1-y^{2i})}~,& k\geq M~,\\
0~, & k<M~.
\end{cases}\label{eqgrassy}
\end{equation}
In particular, evaluating at $y=1,$ we have 
\begin{equation}
\Omega(M,1,k)=\chi(\mathcal{M}^{k}_{M,1})=\begin{cases}\binom{k}{M}~,& k\geq M~, \\
0~, & k<M~.
\end{cases}\label{eqgrass}
\end{equation}  

This result serves two purposes.  First, the explicit formulas \eqref{eqgrassy}-\eqref{eqgrass} serve as useful grounding cases against which we can benchmark our more general results.  Second, it provides some intuition about the nature of Kronecker moduli spaces.  The Grassmannian $Gr(M,k)$ describes vectors in generic position (enforced by the D-term equation \eqref{dterm}).  The Kronecker moduli spaces $\mathcal{M}^{k}_{M,N}$ generalize this idea, and describe configurations of matrices in general position.

\subsubsection{Moduli Spaces as a Function of $k$}

Another way to gain intuition about Kronecker moduli spaces is to examine them as a function of the number of arrows $k.$   Consider the special case $M=N$.  Then, the moduli space describes $k$ linear maps
\begin{equation}
\Phi_{i}: \mathbb{C}^{M}\rightarrow \mathbb{C}^{M}~.
\end{equation}
As a consequence of the D-term equation, these maps are in general position and hence, on an open set in the moduli space, at least one of them is invertible.  One may then fix some of the gauge redundancy by going to a basis where this invertible map is the identity.

What remains after this is $k-1$ linear maps, where now the remaining gauge redundancy acts as conjugation.  For $k=1$ this problem is trivial.  For $k=2,$ this problem is solved by the Jordan decomposition theorem.  Correspondingly, for $k=1,2$ all Kronecker moduli spaces may be explicitly determined.

Finally, for $k>2$ these moduli spaces parameterize multiple linear maps up to conjugation.  This is a notoriously wild representation theory problem, and there is no known general description of the moduli space $\mathcal{M}^{k}_{M,N}$.  Despite the complexity for large $k,$ we will be able to obtain simple exact formulas for the numerical invariants of these moduli spaces.

\subsection{The Degeneration Formula and Star Quivers}
\label{sec:degen}

In order to determine a formula for the Euler characteristic of Kronecker moduli space, it is useful to reduce it to a sum of Euler characteristics of simpler spaces.  This is achieved with the MPS degeneration formula \cite{Manschot:2010qz} (see also \cite{Reineke:2011}).

Physically speaking, the MPS degeneration formula expresses the bound states as a sum of contributions where the $M$ particles of type one group into clumps, each of which subsequently interact and form bound states with the remaining particles of type two.  Weighting each contribution appropriately, we determine that the Euler characteristic of the Kronecker moduli space may be written as
\begin{align}\label{MPS}
\Omega(M,N,k)=\chi(\mathcal{M}^k_{M,N})  = \sum_{m_*\vdash M } \left[\prod_{\ell=1}^M{1\over m_\ell!} \left( (-1)^{\ell-1}\over \ell^2\right)^{m_\ell} \right]\chi ( \mathcal{M}^k_{m_*,N})~, 
\end{align}
where the sum is over all partitions $m_*=(m_1,\cdots,m_M)$ of $M$, 
\begin{align}
\sum_{\ell= 1}^M \ell m_\ell = M,~~~~m_\ell\in \mathbb{N}\cup \{0\}~.
\end{align}

\begin{figure}[here!]
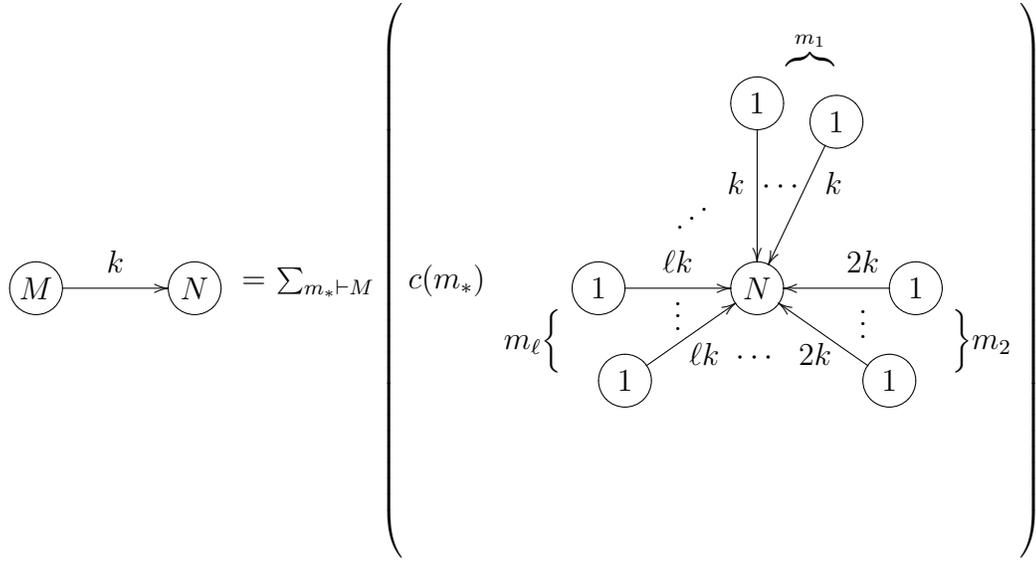

\centering
\subfloat{
\xy  0;<1pt,0pt>:<0pt,-1pt>::
(-300,0) *+{M}*\cir<10pt>{} ="1",
(-240,0) *+{N}*\cir<10pt>{} ="2",
(-270, -10) *+{k} ="b",
\ar @{->} "1"; "2"
\endxy}
~~=~$\sum_{m_*\vdash M}\left( \,c(m_*)
\subfloat{
\xy  0;<1pt,0pt>:<0pt,-1pt>::
(-240,0) *+{N}*\cir<10pt>{} ="1",
(-300,0) *+{1}*\cir<10pt>{} ="2",
(-290,35) *+{1}*\cir<10pt>{} ="3",
(-180,0) *+{1}*\cir<10pt>{} ="6",
(-190,35) *+{1}*\cir<10pt>{} ="7",
(-210,-63) *+{1}*\cir<10pt>{} ="5",
(-240,-70) *+{1}*\cir<10pt>{} ="4",
(-270, -10) *+{\ell k} ="b",
(-265, -30) *+{\iddots} ="b",
(-270, 7) *+{\vdots} ="b",
(-260, 25) *+{\ell k} ="a",
(-325, 20) *+{m_\ell\Big\{} ="a",
(-248, -40) *+{k} ="a",
(-200, -10) *+{2k} ="a",
(-200, 10) *+{\vdots} ="a",
(-155, 20) *+{\Big\}m_2} ="a",
(-230, 25) *+{\cdots~ ~2k} ="a",
(-225, -40) *+{\,\,\cdots ~~ k} ="a",
(-220, -90) *+{\overbrace{}^{m_1}} ="a",
\ar @{->} "2"; "1"
\ar @{->} "3"; "1"
\ar @{->} "4"; "1"
\ar @{->} "6"; "1"
\ar @{->} "5"; "1"
\ar @{->} "7"; "1"
\endxy}
\right)$
  \caption{A graphical representation of the MPS degeneration formula \eqref{MPS}. The Euler characteristic of the Kronecker moduli space can be expressed as a sum of Euler characteristics of star quivers, each associated to a partition of $M$.  The contribution of each star quiver is weighted with a combinatorial coefficient $c(m_*)= \prod_{\ell=1}^M{1\over m_\ell!} \left( (-1)^{\ell-1}\over \ell^2\right)^{m_\ell} .$}
\label{fig:MPS}
\end{figure}

The main actor in the degeneration formula, is the Euler characteristic of the space $\mathcal{M}^k_{m_*,N}.$  This is the quiver moduli space of the ``star quiver" (Figure \ref{fig:MPS}) associated to the partition $m_*$.  It has a central non-abelian node with rank $N$ and $\sum_{\ell=1}^Mm_\ell$ abelian nodes surrounding the non-abelian node. Among all the abelian nodes, $m_\ell$ of them have $\ell k$ arrows pointing to the non-abelian node for each $\ell=1,\cdots,M$ ($m_\ell$ may vanish).  

As a result of the degeneration formula \eqref{MPS}, the task of computing the Euler characteristic of Kronecker moduli space is reduced to determining the Euler characteristics of star quivers.  We carry out this task using the supersymmetric localization formula which expresses the Euler characteristic of quiver moduli spaces as a Jeffrey-Kirwan residue integral \cite{Cordova:2014oxa,Hori:2014tda}.  

In the case of the star quiver associated to the partition $m_*\vdash M$ with general $(M,N)$, the combinatorics of the residues is involved. However in the special case
\begin{align}
N=Mr+1~, \label{Nsimp}
\end{align}
for some non-negative integer $r$, the combinatorics simplifies to an elementary problem in graph theory.  From now on, we restrict our analysis to the special case where \eqref{Nsimp} holds.  In the remainder of this section we describe the resulting correspondence between residues and graphs, and state the solution to the resulting counting problem.  Complete proofs are deferred to \S \ref{sec:derivation} and \S \ref{sec:graph} respectively.

In \S \ref{sec:derivation}, we demonstrate that when $N=Mr+1,$ the Euler characteristic of the star quiver may be expressed as
\begin{align}
\chi(\mathcal{M}_{m_*,Mr+1}^k) = {1\over (Mr+1)!}\,
T(\vec{L}_{m_*,r})
\,
\prod_{\ell= 1}^M
\left [ { \ell k \choose \ell r+1 } (\ell r+1)!\right]^{m_\ell}~.  \label{starcharacter}
\end{align}
One significant feature of the above formula is that the $k$ dependence has been solved.  Meanwhile, the quantity $T(\vec{L}_{m_*,r}),$ is a positive integer depending only on the partition $m_{*},$ and the integer $r$.  As we describe in the next subsection, $T(\vec{L}_{m_*,r})$ has a graph theoretic interpretation which enables us to determine it explicitly as well.

\subsection{The Graph Theory Problem: Counting Trees}\label{subsec:graph}

In this section, we describe the general graph theory problem to which the quantity $T(\vec{L}_{m_*,r})$ is the solution.  The nature of this problem is a direct result of the combinatorics of Jeffrey-Kirwan residues described in \S \ref{sec:derivation}.  A complete treatment of the relevant graph theory (along with proofs omitted here) is given in \S \ref{sec:graph}.  

\begin{figure}[h]
\begin{center}
\includegraphics[width=.6\textwidth]{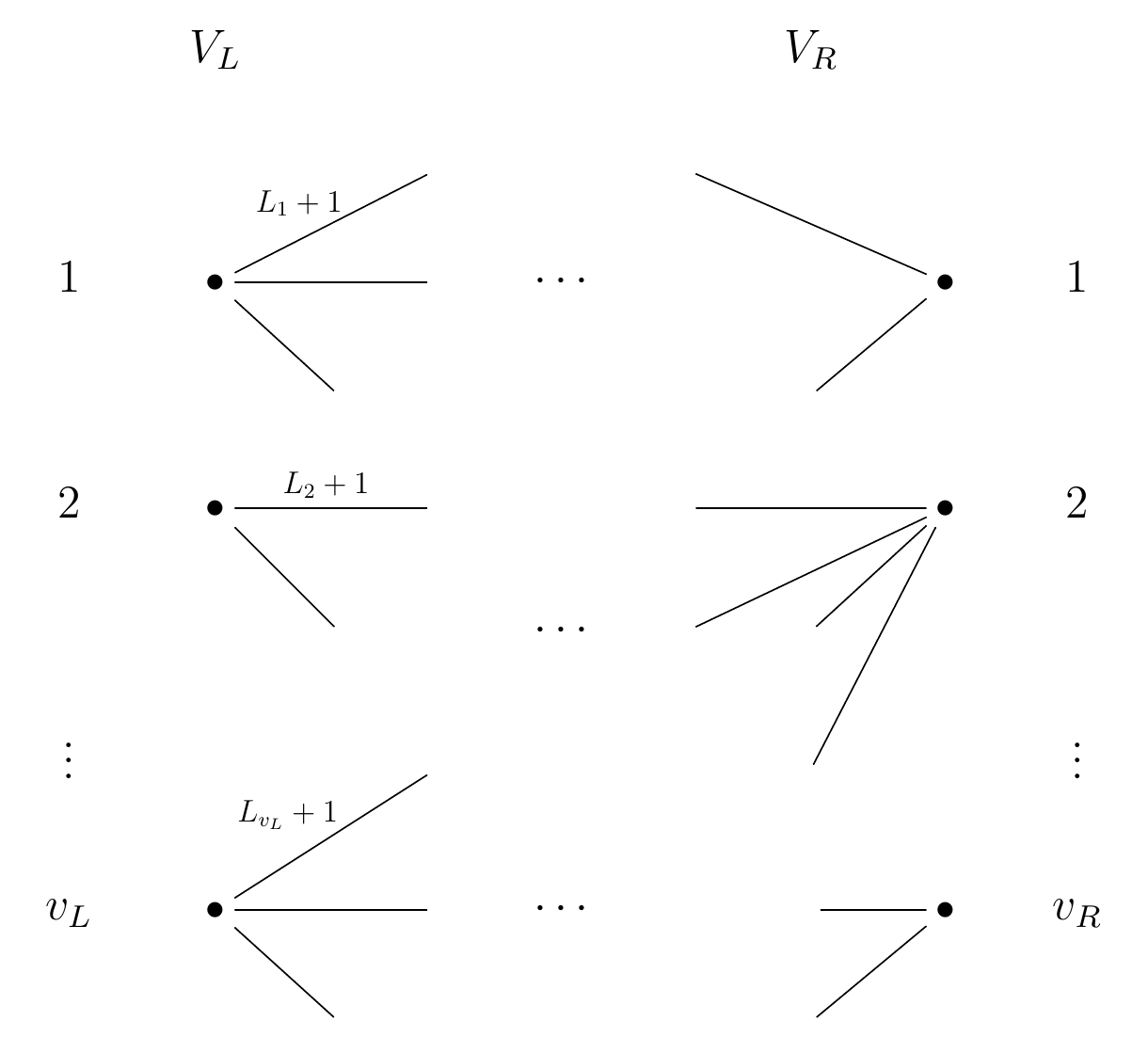}
\end{center}
\caption{A  bipartite graph $G$ in the set $\mathcal{G}(\vec{L})$ defined by a $v_L$-dimensional vector $\vec{L}=(L_1,\cdots,L_{v_L})$ with $L_i \in \mathbb{N}\cup \{0\}$. (The connections between partite sets are not shown.) There are $v_L$ vertices in $V_L$ and $v_R = 1+\sum_i L_i$ vertices in $V_R$. The $i$-th vertex in $V_L$ is incident to $L_i+1$ edges. The number of edges for a vertex in $V_R$ is not constrained. The total number of edges is $e= v_L + \sum_i L_i = v_L+v_R-1$. The quantity of interest $T(\vec{L})$ is the number of trees (connected graphs with no cycle) in the set of bipartite graphs $\mathcal{G}(\vec{L})$.   }\label{fig:graphproblem}
\end{figure}

The graph theory problem of interest concerns counting trees in \textit{bipartite graphs}.   Such a graph $G$ is defined by a pair $G=(V_L+V_R,E)$ with $V_L$ and $V_R$ (the \textit{partite sets}) two disjoint sets of vertices and $E$ the set of edges. The bipartite structure means that edges connect vertices in $V_L$ and $V_R$, but do not connect vertices in the same partite set.   The graphs we consider are undirected, meaning that the edges do not have any orientation.

Now, given a positive integer $v_L,$ and  $\vec{L}=(L_1,L_2,\cdots, L_{v_L})$ a $v_L$-dimensional vector with non-negative integral components, we consider bipartite graphs $G=(V_L+V_R,E)$ satisfying the following incidence relations:
\begin{itemize}
\item There are $v_L$ vertices in $V_L$.
\item There are $v_R=1+\sum_{i=1}^{v_L}L_i$ vertices in $V_R$.
\item There are $L_i+1$ edges incident to the $i$-th vertex in $V_L$.
\end{itemize}
This definition is asymmetric between the partite sets $V_{L}$ and $V_{R}.$  In particular, the graphs in question have an unconstrained number of edges incident at vertices in $V_{R}$.  We denote the set of bipartite graphs satisfying these incidence data by $\mathcal{G}(\vec{L})$.

Define a \emph{tree,} to be a connected graph with no cycles.  The main quantity of interest is then
\begin{align}\label{T(L)}
T(\vec{L} ) = ~\text{number of trees in $\mathcal{G}(\vec{L})$}~.
\end{align}
The incidence data defining the bipartite graphs $\mathcal{G}(\vec{L})$ are tuned in such a way as to make the number of trees non-zero.  Indeed, using $v_R= 1+\sum_i L_i$, the total number of edges $e$ is related to the total number of vertices $v=v_L+v_R$ by
\begin{align}
e = \sum_{i=1}^{v_L} (1+L_i) = v_L +v_R-1 = v-1~.
\end{align}
By an elementary proposition in graph theory (see Proposition \ref{prop:tree} in \S \ref{sec:graph}), a tree has precisely $v-1$ edges. Hence it is precisely for this choice of $v_R$ that the set of trees can be non-empty.  We illustrate these definitions in several examples below.

\paragraph{Example:  $v_L=1,$  $\vec{L}=(L_1)$}\mbox{}\\

\begin{figure}[h]
\begin{center}
\includegraphics[width=.4\textwidth]{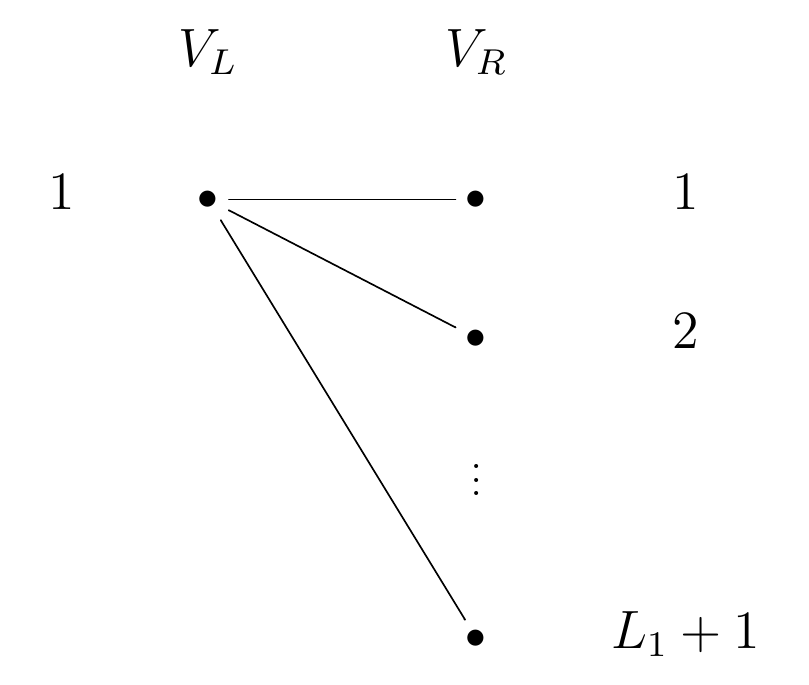}
\end{center}
\caption{An example of a tree in $\mathcal{G}(\vec{L})$ with $\vec{L}= (L_1)$. We have $v_L=1$ vertex in $V_L$ and $v_R=L_1+1$ in $V_R$, with a total number of $e=L_1+1$ edges. The vertex in $V_L$ is incident to $L_1+1$ edges. This is the unique tree  in $\mathcal{G}(\vec{L})$, \textit{i.e.} $T(\vec{L})=1$. }\label{fig:vL=1}
\end{figure}

The simplest example has a single left vertex  ($v_L=1$) and $\vec{L}=(L_1)$ a one-dimensional vector for some non-negative integer $L_1.$ In this case $v_R=L_1+1$ and the number of edges $e=L_1+1$. For $G\in \mathcal{G}(\vec{L})$, there are $L_1+1$ edges incident to the vertex in $V_L$. There is a unique tree in $\mathcal{G}(\vec{L})$ shown in Figure \ref{fig:vL=1}.

\paragraph{Example: $v_L=2,$  $\vec{L}=(1,1)$}\mbox{}\\

\begin{figure}[h]
\begin{center}

\subfloat[]{\label{fig:6graphs}
{
\includegraphics[width=1\textwidth]{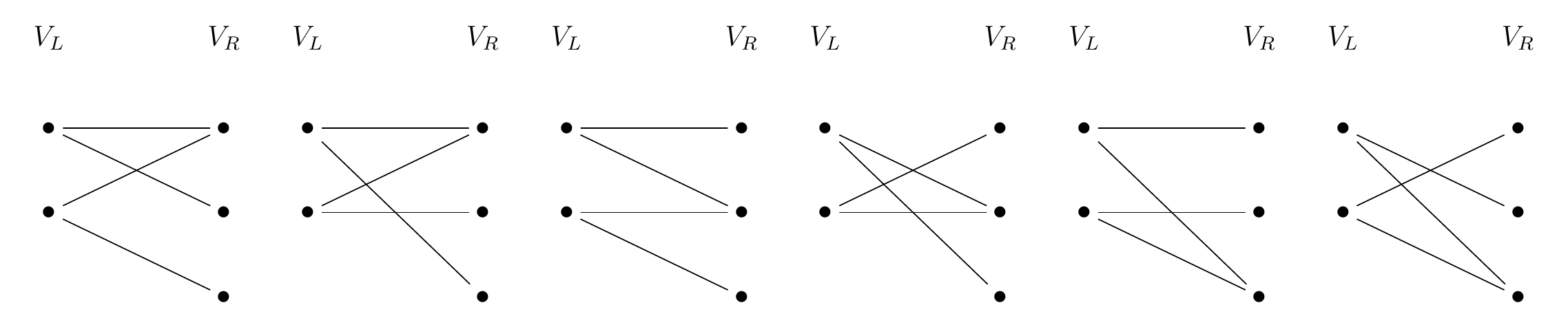}
}
}\\
~\\
~
\subfloat[]{\label{fig:cyclic}
{
\includegraphics[width=.3\textwidth]{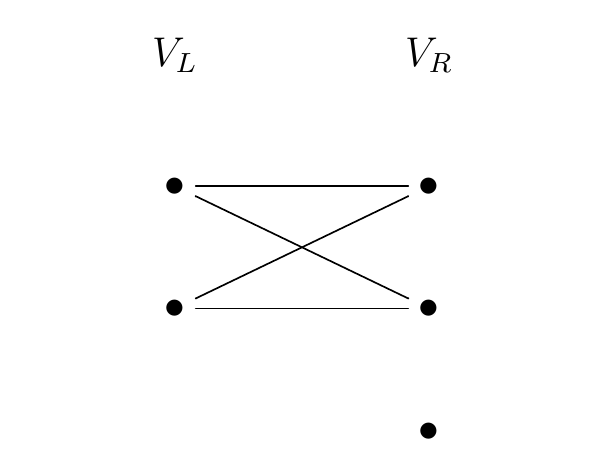}
}
}
\end{center}
\caption{(a) The 6 trees in $\mathcal{G}(\vec{L})$ with $\vec{L}= (1,1)$.  (b) A graph in $\mathcal{G}(\vec{L})$ that is not a tree.}
\end{figure}

As a more advanced example, consider the case with two left vertices ($v_L=2$) and $\vec{L}=(1,1)$ in Figure \ref{fig:6graphs}. There are three right vertices  ($v_R=3$) and four total edges ( $e=4$). Each of the vertices in $V_L$ is incident to 2 edges. There are 6 trees in $\mathcal{G}(\vec{L})$ shown in Figure \ref{fig:6graphs}, \textit{i.e.} $T(\vec{L})=6$.

\subsubsection{The Tree Counting Theorem and $T(\vec{L}_{m_*,r})$}
\label{subsec:tree}

One of the remarkable facts about trees is that given any incidence data defining the set of bipartite graphs $\mathcal{G}(\vec{L}),$ the number of trees $T(\vec{L})$ can be exactly determined.  The result is a key theorem:\footnote{We thank Fan Wei and Yu-Wei Fan for pointing out that this theorem appears in the textbook \cite{Stanley}.}
\begin{theorem}\label{theorem}
The number of trees in the set $\mathcal{G}(\vec{L})$ is
\begin{align}\label{mainthm}
T(\vec{L}) = {(v_R-1)!\over \prod_{i=1}^{v_L}L_i!}v_R^{v_L-1}, 
\end{align}
where $v_R=1+\sum_{i=1}^{v_L}L_i$.
\end{theorem}

In \S \ref{sec:graph} we develop the necessary graph theory to prove \eqref{mainthm} directly. For now, we apply this result to the problem of the Euler characteristics of star quivers.  To do so, it remains to identify the quantity $T(\vec{L}_{m_*,r})$ of \eqref{starcharacter}.  This is a special case of the tree function $T(\vec{L})$  defined in \eqref{T(L)}.

Given a partition $m_* = (m_1,\cdots,m_M)$ of $M$, \textit{i.e.} $m_\ell\ge0$ and $\sum_\ell \ell m_\ell=M$, and a non-negative integer $r$, we associate the following choice of the vector $\vec{L}$:
\begin{align}\label{Lmj}
\vec{L}_{m_*,r} = \left(  \underbrace{r,\cdots,r}_{m_1},   \cdots,  \underbrace{\ell r,\cdots,\ell r}_{m_\ell},\cdots, \underbrace{Mr,\cdots,Mr}_{m_M}\right)~.
\end{align}
The number of vertices and edges in such trees are read off from the above.  We have
\begin{equation}
v_L = \sum_{\ell=1}^{M} m_\ell~, \hspace{.5in}v_R =Mr+1~, \hspace{.5in} e= Mr +\sum_{\ell=1}^M m_\ell~.
\end{equation}
These quantities are related to the physical parameters defining the star quiver of Figure \ref{fig:MPS}.  In particular, $v_{L}$ is the number of abelian nodes, while $v_{R}$ is the rank of the central non-abelian node.  Finally, the structure of  $\vec{L}_{m_*,r} $ emerges from the pattern of Fayet-Iliopoulos parameters as derived in \S \ref{sec:derivation}.

Applying the tree counting theorem \ref{theorem} we conclude that the number of bipartite trees with this incidence data is given by
\begin{align}\label{Tmj}
 T(\vec{L}_{m_*,r})={(Mr)!\over \prod_{\ell=1}^M(\ell r)!^{m_\ell} } (Mr+1)^{\sum_\ell m_\ell -1}~.
\end{align}

\subsection{An Index Formula for the Kronecker Quiver $(M,Mr+1)$}
\label{subsec:result}

Armed with an explicit expression for $T(\vec{L}_{m_*,r})$ and the factorization formula \eqref{starcharacter}, we are now ready to present our main result. The index for the Kronecker quiver quantum mechanics with dimension vector $(M,Mr+1)$ and $k$ arrows is: 
\begin{align}\label{final2}
\chi(\mathcal{M}^k_{M,Mr+1}) ={1\over (Mr+1)^2} \sum_{m_*\vdash M}
\prod_{\ell=1}^M  {1\over m_\ell!} \left[ 
{(-1)^{\ell -1}\over \ell^2}  { \ell k\choose \ell r +1 } (Mr+1)(\ell r +1)
\right]^{m_\ell}~,
\end{align}
where the sum is over the partitions $m_*=(m_1,\,m_2,\,\cdots,\, m_M)$ of $M$, \textit{i.e.} $\sum_{\ell} \ell m_\ell =M$ and $m_\ell \in \mathbb{N}\cup \{0\}$.

The expression \eqref{final2} may be simplified with the help of an exponential resummation and the introduction of a generating function.  Introduce the function  
\begin{align}\label{Fkjx}
F(k,r,x) =(k-r) \sum_{\ell=1}^\infty  {(-1)^{\ell-1}\over \ell} {k\ell \choose r\ell }  x^\ell~,
\end{align}
and let $[x^j]\{ q(x) \}$ denote the coefficient of $x^{j}$ in a power series  $q(x)$.  Then, using the identity 
\begin{align}\label{Id}
\sum_{M=0}^\infty\sum_{m_*\vdash M} \prod_{\ell=1}^M {1\over m_\ell!}\left( p(\ell) x^\ell \right)^{m_\ell}
= \exp \left[ \sum_{\ell=1}^\infty p(\ell) x^\ell\right]~,
\end{align}
we can express \eqref{final2} as
\begin{align}\label{final3}
\boxed{\,
\Omega(M,Mr+1,k)=\chi(\mathcal{M}^k_{M,Mr+1}) = {1\over (Mr+1)^2}
[x^M] \left\{
\exp\Big[\, (Mr+1) F(k,r,x)\, \Big] \right\}  ~.\,}
\end{align}
This is our final result for the general Euler characteristic of Kronecker moduli space.

\section{Analysis of the Index Formula}
\label{sec:check}

In this section, we provide a number of checks on the index formula.  

In \S \ref{sec:special} we illustrate simplifications that occur in the index formula at special values of $r.$  Specifically, we recover the Grassmannian index at $r=0$ and determine a closed form expression for the degeneracies when $r=1.$  In \S \ref{sec:sym} we demonstrate that the index formula satisfies required symmetry properties arising from isomorphisms of Kronecker moduli spaces.  Finally, in \S \ref{sec:wall}, we directly compare our index result \eqref{final3} to wall-crossing formulas. 

\subsection{Simplifications at Special Values of $r$}
\label{sec:special}

In this section we describe simplifications to the Euler characteristic formula which occur at special values of $r$.

\subsubsection{Recovering the Grassmannian Index}

The first check on our result occurs when $r=0.$  In that case the dimension vector is $(M,1)$ and hence as reviewed in \S \ref{sec:grass}, the moduli space is a Grassmannian.  Thus, for $r=0$ we must recover $\chi(Gr(M,k))$.  This is readily verified.  Indeed, when $r=0,$ the function $F(k,r=0,x)$ of \eqref{Fkjx} simplifies to 
\begin{align}
F(k,r=0,x) = k  \sum_{\ell=1}^\infty  {(-1)^{\ell-1}\over\ell} x^\ell = k \log(1+x)~.
\end{align}
The Euler characteristic of the quiver moduli space $\mathcal{M}^k_{M,1}$ is then determined from the prescription \eqref{final3} as
\begin{equation}
\chi(\mathcal{M}^k_{M,1})= [x^M]\Big\{ \exp\left[ F(k,0,x)\right] \Big\}=[x^M] \left\{ (1+x)^k\right\} = { k\choose M}~,
\end{equation}
which is indeed the Euler characteristic for the Grassmannian $Gr(M,k)$.

\subsubsection{The Case $r=1$}

The index formula \eqref{final3} also simplifies in the special case $r=1$ (as well as the equivalent case $r=k-1$ via \eqref{iso2}).  

In that case, the function $F(k,1,x)$ of \eqref{Fkjx} may be rewritten in terms of the \emph{generalized binomial series} $\mathcal{B}_{k}(x)$ defined as \cite{Graham:1994:CMF:562056}
\begin{equation}
\mathcal{B}_{k}(x)=\sum_{\ell=0}^{\infty} {k \ell +1 \choose \ell}\frac{x^{\ell}}{k\ell+1}~.
\end{equation}
The generalized binomial series obeys an algebraic identity
\begin{equation}
\mathcal{B}_{k}(x)=1+x \mathcal{B}_{k}(x)^{k}~,
\end{equation}
which enables one to easily determine the power series coefficients of $\mathcal{B}_{k}(x)$ raised to an arbitrary real power
\begin{equation}
[x^m] \left\{ \mathcal{B}_{k}(x)^{s}\right\} = {mk+s\choose m}\frac{s}{mk+s}~.
\end{equation}
In particular, in the $s\rightarrow0$ limit we then obtain
\begin{equation}
\log\left(\mathcal{B}_{k}(x)\right)=\sum_{\ell=1}^{\infty}{k \ell \choose \ell}\frac{x^{\ell}}{k\ell}~.
\end{equation} 
From which we deduce that
\begin{equation}
\exp[F(k,1,x)]=\mathcal{B}_{k}(-x)^{k(1-k)}~. \label{binomialseries}
\end{equation}

Comparison to the index formula \eqref{final3}, now yields the following simple expression for the index
\begin{equation}
\Omega(M,M+1,k)=\chi(\mathcal{M}^k_{M,M+1})  ={k\over (M+1) \left[(k-1)M +k\right] }{ (k-1)^2 M +k(k-1) \choose M}~. \label{weist}
\end{equation}
The result \eqref{weist} was first obtained by  Weist \cite{Weist:2009} by alternative techniques.  The fact that we have reproduced \eqref{weist} using the combinatorics of Jeffrey-Kirwan residues provides a strong check on our method. 

Beyond the direct check provided on our work, the appearance of the generalized binomial series is in and of itself significant.  This function, $\mathcal{B}_{k}(x),$ has a natural combinatorial interpretation as the generating function of rooted $k$-ary trees.  These binomials series also make an appearance in the generating function of indices $\Omega(M,M,k)$ \cite{Kontsevich:2008fj, Gross, Galakhov:2013oja}.  In that case, since the dimension vector $(M,M)$ is not coprime, the quiver quantum mechanics has bound states at threshold.  The associated Kronecker quiver moduli spaces have singularities, and extracting the indices from geometry is delicate.  Nevertheless, the resulting generating function again involves the generalized binomial series and suggests a direct interpretation of the index and bound states in terms of rooted trees, perhaps as attractor flow trees \cite{Denef:2000nb, Denef:2000ar}.

\subsection{Symmetries of the Index Formula}
\label{sec:sym}

As a further consistency check, in this section we show our formula \eqref{final3} respects the isomorphisms of the quiver moduli space \eqref{mutation}:
\begin{align}
\begin{split}
&\Omega(M,N,k)= \Omega(N,M,k)~, \\
&\Omega(M,N,k)= \Omega(N,Nk-M,k)~. 
\end{split}
\end{align}
Among all the isomorphisms, the following two infinite families are of particular interest. 
\begin{align}\label{iso1}
\Omega (2,\, 2r+1,\,k) = \Omega ( 2, \, 2(k-r-1)+1,\,k )~,
\end{align}
and
\begin{align}\label{iso2}
\Omega(M,M+1,k)=\Omega\left( M+1, \, (k-1)(M+1)+1,\,k\right)~.
\end{align}
This is because the dimension vectors on both sides of the isomorphisms are of the special case $N=Mr+1$, for which our formula \eqref{final3} is applicable. We will explicitly prove the invariance of our formula \eqref{final3} in the following.

\paragraph{The First Isomorphism}\mbox{}\\

For the first isomorphism \eqref{iso1}, it is easier to use the alternative expression \eqref{final2} for our formula for the index. Using \eqref{final2}, the left-hand-side of \eqref{iso1} becomes
\begin{align}\label{LHSiso1}
\Omega(2,2r+1,k) = -{1\over 4} { 2k \choose 2r+1} +{1\over 2} \left[ {k!\over r! (k-r-1)!}\right]^2.~
\end{align}
Similarly, the righthand side of \eqref{iso1} may be written as
\begin{align}
\Omega(2,\,2(k-r-1)+1,k) =-{1\over 4} {2k\choose 2k-2r-1} +{1\over 2} \left[ {k! \choose (k-r-1)!r!}\right]^2~,
\end{align}\
which is manifestly the same as \eqref{LHSiso1}.

\paragraph{The Second Isomorphism}\mbox{}\\

The second isomorphism \eqref{iso2} is harder to prove. First let us note the following identity from the definition of $F(k,r,x)$ \eqref{Fkjx},
\begin{align}\label{Fkk1}
F(k,\, k-1,x) = {1\over k-1} F(k,1,x)~.
\end{align}
Applying our general formula \eqref{final3} to the righthand side of \eqref{iso2} , we have
\begin{align}\label{RHSiso2}
\begin{split}
\Omega&\left(M+1, \, (k-1)(M+1)+1,\,k\right)\\
&={1\over \left[ (k-1)(M+1) +1\right]^2}
[x^{M+1}] \left\{
\exp \left[ \left( M+1+{1\over k-1}\right) F(k,1,x)\right]
\right\}~,
\end{split}
\end{align}
where we have used the identity \eqref{Fkk1}.

The relationship between $F$ and the generalized binomial series in the case $r=1$ stated in equation \eqref{binomialseries} (as well as the direct argument given in Appendix \ref{sec:powerseries} ) shows that function $F(k,r,x)$ satisfies the following identity
\begin{align}\label{identity}
[x^M] \Big\{ \exp \left[ \,\beta F(k,1,x) \,\right] \Big\}  =  {\beta\over M} k(k-1){ k(k-1)\beta-(k-1)M-1\choose M-1}~,
\end{align}
for any complex number $\beta$. Applying \eqref{identity} with $\beta= M+1+{1\over k-1}$ to \eqref{RHSiso2}, we have
\begin{eqnarray}
\Omega\left(M+1, \, (k-1)(M+1)+1,\,k\right) & = & {1\over (k-1)M+k} \,{k\over M+1} { (k-1)^2 M +k(k-1)\choose M} \nonumber \\
& = &\Omega(M,M+1,k)~,
\end{eqnarray}
where in the last equality we have used the simplified expression \eqref{weist} in the case $r=1$. Hence we have checked that our formula \eqref{final3} respects the isomorphisms \eqref{mutation}.

\subsection{Comparison to Wall-Crossing Formulas}
\label{sec:wall}

A final check on our results is to directly compare the output to wall-crossing formulas.  This allows us to check our result \eqref{final3} for general $r$ and for $M$ and $k$ sufficiently small.

The wall-crossing formula of \cite{Kontsevich:2008fj} allows us to determine the change in the indices $\Omega(M,N,k)$ as the Fayet-Iliopoulos parameters $\zeta$ are varied.  Mathematically, this is the change in the Donaldson-Thomas invariants of the quiver, as the stability condition is changed.

In the context of the Kronecker model, the wall-crossing formula is particularly simple to apply.  By changing the sign of the FI parameter $\zeta$ of \eqref{dterm}, we reach a chamber where all moduli spaces are empty, and no non-trivial bound states can form.  The only stable states are then a single particle of type one, or a single particle of type two.  We may use this simple chamber ($\zeta<0$) as a seed and apply the wall-crossing formula to determine the indices in the chamber of interest ($\zeta>0$).

To phrase the wall-crossing computation we first introduce functions $K_{M,N}$ which act on formal variables $\Big[x,y\Big]$ as
\begin{equation}
K_{M,N}\Big[  x,y \Big]= \Big[\, x(1-(-1)^{kMN}x^{M}y^{N})^{kN}, y(1-(-1)^{kMN}x^{M}y^{N})^{-kM}\,\Big]~.
\end{equation}
We also introduce a sign function which detects the parity of the dimension of $\mathcal{M}^{k}_{M,N}$
\begin{equation}
\sigma(M,N,k)=\begin{cases}+1~, &  kMN-M^{2}-N^{2}+1 \equiv 0 \  (\mathrm{mod} \ 2)~,\\
-1~, & kMN-M^{2}-N^{2}+1\equiv 1 \  (\mathrm{mod} \ 2)~.\end{cases}
\end{equation}

The wall-crossing formula asserts that a particular function of $[x,y]$ built via compositions of the above functions does not depend on the chamber.  In the case at hand this implies that 
\begin{equation}
\prod_{M,N \geq 0}^{\rightarrow}K_{M,N}^{\sigma(M,N,k)\Omega(M,N,k)} = K_{0,1} \circ K_{1,0}~, \label{wallcrossing}
\end{equation}
Where in the above, the product operation is composition of functions, and the order of composition is that of decreasing $M/N.$\footnote{If $M_{1}/N_{1}=M_{1}/N_{2}$ then $K_{M_{1},N_{1}}\circ K_{M_{2},N_{2}}=K_{M_{2},N_{2}}\circ K_{M_{1},N_{1}}.$  Also the need to introduce $\sigma$ is because we have defined the index $\Omega$ to coincide with the Euler characteristic.}

To use the wall-crossing formula, note that the function $K_{M,N}$ differs from the identity only at order $x^{M}y^{N}.$  Whence if we fix a positive integer $Q$ we may solve \eqref{wallcrossing} order by order by replacing the infinite product with a finite product where only those $K_{M,N}$ are retained with $M+N\leq Q$.  Then we may compute the composition as a formal power series (again only retaining terms differing from the identity up to total order $Q$).  Matching to the right-hand side, we then solve for all $\Omega(M,N,k)$ with $M+N\leq Q$.

This procedure is computationally costly to carry out for large $Q$.  However, for $Q$ sufficiently small, wall-crossing provides a useful crosscheck on our results in the case where $r\neq 0,1.$  As an example in Table \ref{tab:data}, we present wall-crossing results for $r=2$.  All indices computed thus far agree with our general index prescription \eqref{final3}, and provides a strong check on the validity of our result.

\renewcommand{\arraystretch}{1.5}
\begin{table}[H]
\small
  \centering
  \begin{tabular}{ |c | c | c|c|}
\hline
{\bf r=2} & {$k=4$} & {$k=5$}& {$k=6$} \\
\hline
\hline
{$M=0$}& {$1$} &1&1\\
\hline
{$M=1$}& {$4$} &10&20\\
\hline
{$M=2$}& {$58$} &387&1602\\
\hline
{$M=3$}& {$1264$}&22765& 196136\\
\hline
{$M=4$}& {$33751$}&1647265&29595451\\
\hline
{$M=5$}& {$1018252$}&134924271&5059514952\\
\hline
{$M=6$}& {$33331794$}&12003007130&939887566862\\
\hline
{$M=7$}& {$1156714720$}&1132713788250&185267731189312\\
\hline
{$M=8$}& {$41942224979$}&111732981265605&38180107620131049\\
\hline
{$M=9$}& {$1573700333920$}&11407942652134355&8145089457477277400\\
\hline
{$M=10$}& {$60685423064290$}&1197321303724493265&1786374698749585024792\\
\hline
{$M=11$}& {$2393245242889696$}&128534027591027982405&400759043478342386735448\\
\hline
{$M=12$}& {$96164522270610418$}&14060437197335739888902&91619999838823992655697998\\
\hline
{$M=13$}& {$3925754745132237556$}&1562780288066103516885980& 21283030090887836618088693536\\
\hline
\end{tabular}
  \caption{Indices $\Omega(M,Mr+1,k),$ extracted from the wall-crossing formula for  $r=2$ and various values of $k$ and $M.$  The data agrees with our general index formula. }
  \label{tab:data}
\end{table}

\section{Bipartite Graphs from Contour Integrals}\label{sec:derivation}

\begin{figure}[here!]
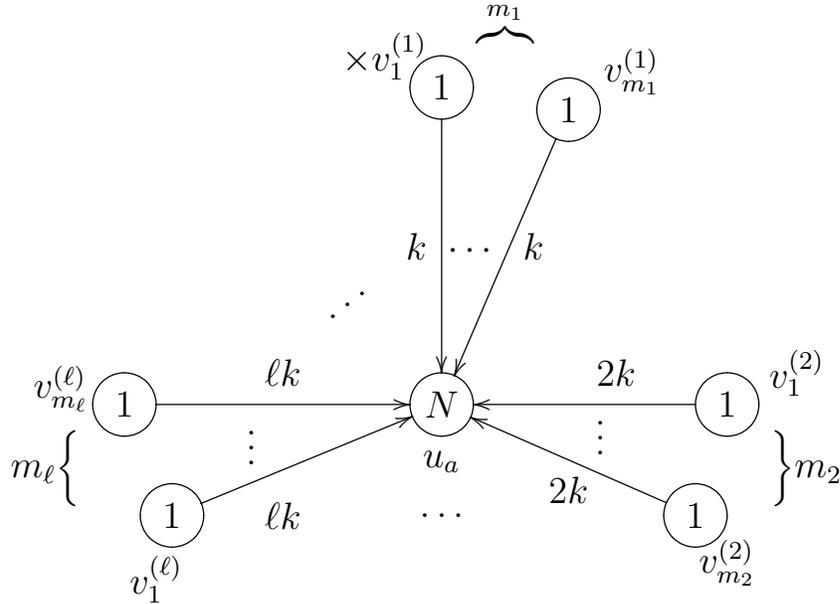

  \centering
  \scalebox{1.2}{
\subfloat{
\xy  0;<1pt,0pt>:<0pt,-1pt>::
(-240,0) *+{N}*\cir<10pt>{} ="1",
(-340,0) *+{1}*\cir<10pt>{} ="2",
(-325,35) *+{1}*\cir<10pt>{} ="3",
(-150,0) *+{1}*\cir<10pt>{} ="6",
(-160,35) *+{1}*\cir<10pt>{} ="7",
(-200,-93) *+{1}*\cir<10pt>{} ="5",
(-240,-100) *+{1}*\cir<10pt>{} ="4",
(-290, -10) *+{\ell k} ="b",
(-270, -35) *+{\iddots} ="b",
(-190, 7) *+{\vdots} ="b",
(-290, 35) *+{\ell k} ="a",
(-365, 20) *+{m_\ell\Big\{} ="a",
(-248, -50) *+{k} ="a",
(-185, -10) *+{2k} ="a",
(-300, 10) *+{\vdots} ="a",
(-125, 20) *+{\Big\}m_2} ="a",
(-200, 27) *+{2k} ="a",
(-225, -50) *+{\,\,\cdots ~~ k} ="a",
(-220, -120) *+{ \overbrace{}^{m_1} } ="a",
(-240, 35) *+{\dots} ="a",
(-258, -110) *+{\times v_1^{(1)} } ="a",
(-180, -105) *+{ v_{m_1}^{(1)} } ="a",
(-150, 50) *+{ v_{m_2}^{(2)} } ="a",
(-128, -10) *+{ v_1^{(2)} } ="a",
(-330, 55) *+{ v_{1}^{(\ell)} } ="a",
(-360, -5) *+{ v_{m_\ell}^{(\ell)} } ="a",
(-240, 18) *+{ u_a } ="a",
\ar @{->} "2"; "1"
\ar @{->} "3"; "1"
\ar @{->} "4"; "1"
\ar @{->} "6"; "1"
\ar @{->} "5"; "1"
\ar @{->} "7"; "1"
\endxy}}
  \caption{The ``star quiver" associated to the partition $m_*=(m_1,\cdots,m_M)$ of $M$. There is a single non-abelian node with rank $N$ and $\sum_{\ell=1}^Mm_\ell$ abelian nodes surrounding the non-abelian node. Among all the abelian nodes, $m_\ell$ of them have $\ell k$ arrows pointing to the non-abelian node for each $\ell=1,\cdots,M$ ($m_\ell$ could be zero). The gauge fugacities of the non-abelian node are denoted by $u_a$, $a=1,\cdots, N$, while those of the abelian nodes are denoted by $v^{(\ell)}_{I}$ with $\ell=1,\cdots, M$ and $I=1,\cdots,m_\ell$. The cross $\times$ for $v_1^{(1)}$ means that abelian node is decoupled when we apply the residue formula.}\label{fig:star}
\end{figure}

In this section we derive the graph theory algorithm to compute the refined index of star quivers illustrated in Figure \ref{fig:star} using a Jeffrey-Kirwan residue formula.  

We start in \S \ref{sec:residue} with a presentation of the residue formula for the index of star quiver as derived by supersymmetric localization.  In \S \ref{sec:JKrule} we begin our task of enumerating the poles that contribute to the residue formula.  In \S \ref{sec:polegraph} we demonstrate that every contributing pole corresponds to a choice of bipartite graph.  Finally, in \S \ref{sec:graphpole} we prove that each graph associated to a pole in the index is a tree described by the incidence data used in \S\ref{subsec:graph}.  In particular we derive the Euler characteristic of the star quiver \eqref{stareulerform}.

\subsection{A Residue for the Star Quiver}
\label{sec:residue}
In this section, we introduce the residue formula for the refined index of the star quiver.

We begin with a recollection of the data of the star quiver and introduce the relevant notation.  

Let $e$ denote the total rank of the gauge group of the star quiver.\footnote{As we shall demonstrate, $e$ is also the number of edges in the bipartite trees relevant to the residue, hence the choice of notation.}  The residue formula that we use to compute cohomology of star quivers has $e$ integration variables (known as gauge fugacities).  We label them as follows.
\begin{itemize}
\item The gauge fugacities of the central non-abelian node of the star quiver are indicated by $u_a$, $a=1,\cdots, N.$
\item There are $m_{\ell}$ abelian nodes in the star quiver which each has $\ell k$ arrows to the central non-abelian node. Thus we may label the abelian nodes by a pair of integers $(I,\ell)$ with $\ell=1,\cdots, M$ and $I=1,\cdots,m_\ell.$  Correspondingly, we indicate the gauge fugacity of an abelian node by $v^{(\ell)}_I$.
\end{itemize}

Next, consider the node $(I,\ell)$.  This node has $\ell k$ identical arrows emanating from it.  As a result, the quantum mechanics has an $su(\ell k)$ flavor symmetry rotating these chiral multiplets.  We may introduce fixed flavor fugacities into the contour integral.  
\begin{itemize}
\item The flavor fugacities associated to the $su(\ell k)$ flavor symmetry acting on nodes emanating from the node $(I,\ell),$ are denoted by $\xi_i^{(I,\ell)}$ with $i=1,\cdots,\ell k.$  We choose these flavor fugacites to be generic complex numbers. In particular, this means that the $\xi_i^{(I,\ell)}$ are linearly independent over the rational numbers. Thus, any relation of the form
\begin{equation}\label{rationalrel}
\sum q_{i}^{(I,\ell)}\xi_i^{(I,\ell)}=0~,
\end{equation}
with rational $q_{i}^{(I,\ell)}$ implies that all $q_{i}^{(I,\ell)}$ vanish.
\end{itemize}
It is important that the the refined index $\Omega(\mathcal{M}^k_{m_*,N},y)$ in fact does not depend on the the flavor fugacities due to the $\mathcal{N}=4$ supersymmetry \cite{Hori:2014tda}.  Our purpose in introducing these variables is to separate multiple poles in contour integrals and turn the residue calculation into a combinatorics problem.

Finally, before introducing the residue integral, we recall an important technical point.  To compute the refined index of the quiver quantum mechanics, we must decouple a $u(1)$ vector multiplet; otherwise the fermionic zero modes of the overall $u(1)$ would make the answer zero.\footnote{This is due to the fact that all mater transforms in adjoints and bifundamentals, thus an overall $u(1)$ in the gauge group, decouples. } The index does not depend on which $u(1)$ we decouple. For concreteness, we will decouple the $(I=1,\ell=1)$-th abelian node whose gauge fugacity is $v^{(1)}_1$.\footnote{For partitions with vanishing $m_{1},$ one must decouple an alternate choice of gauge fugacity $v_{I}^{(\ell)}$.  As we will ultimately see, the symmetry between the nodes, including the decoupled one, is restored. Thus, the choice of which node one decouples is irrelevant.}  In the following, $v^{(1)}_1$ is interpreted to vanish when it appears in equations.

We are now prepared to introduce the residue formula for the index.  According to the general supersymmetric localization arguments of \cite{Hwang:2014uwa, Cordova:2014oxa,Hori:2014tda} the index may be expressed as\footnote{We have suppressed the dependence of $\Omega$ on the FI parameter $\zeta$ since we are only interested in the choice \eqref{FI}.}
\begin{align}
\label{JKform}
\Omega(\mathcal{M}_{m_*,N}^k,y) = \left({1\over \sin(z)}\right)^{e}
\sum_{(u_*,v_*)\in \mathfrak{M}^*_{\text{sing}}}
\underset{(u,v)=(u_*,v_*)}{\text{JK-Res}}(\mathbf{Q}(u_*,v_*),\zeta) Z_{\text{1-loop}}(u,v,z,\xi)~,
\end{align}
where $y=e^{iz},$ and $e$ is the  rank of the star quiver after decoupling an overall $U(1):$
\begin{align}
e=N+\sum_{\ell=1}^M m_\ell -1~.
\end{align}
 
 Let us describe in detail the elements of the residue formula \eqref{JKform}.  

\begin{itemize}
 
\item  The one-loop determinant $Z_{\text{1-loop}}$ is\footnote{We use a slightly different conventions on $u$ and $z$ than in \cite{Cordova:2014oxa}. $u_{\text{here}} = \pi u_{\text{there}}$ and $z_{\text{here}}= -\pi z_{\text{there}}$.}
 \begin{align}\label{Z1loop}
 Z_{\text{1-loop}} (u,v,z,\xi)= {1\over N!}\prod_{\substack{a,b=1\\a\neq b}}^N {\sin (u_a-u_b  ) \over\sin(u_a-u_b+z)}
 \prod_{\ell=1}^M \prod_{I=1}^{m_\ell}\prod_{i=1}^{\ell k}
 {\sin(u_a - v_I^{(\ell)} -\xi^{(I,\ell)}_i+z)\over  \sin(u_a - v_I^{(\ell)} -\xi^{(I,\ell)}_i)}~,
 \end{align}
  
 \item The FI parameter  $\zeta \in \mathbb{C}^{e}$  is determined by the MPS formula \cite{Manschot:2010qz} to be
\begin{align}\label{FI}
\begin{split}
\zeta &= \left(\underbrace{{1},\cdots, {1}}_{N} , \,
\underbrace{ - {1\over M} N, \cdots, -{1\over M}N}_{m_1-1},\,\underbrace{ - {2\over M} N, \cdots, -{2\over M}N}_{m_2}, \,\cdots,
\underbrace{ - {\ell\over M} N, \cdots, -{\ell\over M}N}_{m_\ell}, \,\cdots
\right)~,
\end{split}
\end{align}
where we order the variables as $(u_a,v^{(1)}_I, v^{(2)}_I,\cdots, v^{(\ell)}_I,\cdots, v^{(M)}_I)$. Again, we only have $m_1-1$ components equal to $-1/M$ because one of the abelian node  with $k$ arrows is decoupled. Notice that if we had not decoupled the $u(1)$, the sum of the components of $\zeta$ would be $N\times {1\over N} - \sum_\ell m_\ell\times{\ell\over M} = 0$, indicating that the overall $u(1)$ decouples.

\item The summation is over points $(u_*,v_*)$ in the gauge fugacity space where at least $e$ of the sine factors appearing in the denominator of \eqref{Z1loop} vanish.  These are poles which may contribute to the residue extracted by the operator $\text{JK-Res}.$  We explain this operator in detail in the following subsection.

\end{itemize}

\subsection{The Jeffrey-Kirwan Rule}\label{sec:JKrule}

In this section, we describe the residue operator $\text{JK-Res}$ which specifies the contribution of a given pole point $(u_*,v_*)$ in the gauge fugacity space.  We refer to the resulting rule dictating which poles contribute to the index as the ``Jeffrey-Kirwan rule".

Consider a given polar point $(u_*,v_*)$.  This pole is said to be \emph{non-degenerate} if there are exactly $e$ denominator factors in \eqref{Z1loop} which vanish at $(u_*,v_*).$  Meanwhile, if more than $e$  denominator factors in \eqref{Z1loop} vanish at $(u_*,v_*)$ the pole is said to be \emph{degenerate}.  

For general $M,N$, both the degenerate and non-degenerate poles contribute to the index. However, in the case
\begin{align}
N=Mr+1~, \label{Nsimpr}
\end{align}
for some non-negative integer $r$, only the non-degenerate poles contribute. We argue for this significant technical simplification in Appendix \ref{sec:degenerate}.  In the remainder of our analysis, we restrict to the case where \eqref{Nsimpr} holds.  Our task is thus reduced to describing the non-degenerate poles.

At each non-degenerate pole $(u_*,v_*),$ there are exactly $e$ denominator factors that vanish at $(u_*,v_*).$  Each such denominator factor is a sine function whose argument is a linear function of the gauge and flavor fugacities.   The dependence on the gauge fugacities is conveniently encoded by a vector $Q$ composed of the coefficients of the gauge fugacities in the arguments of the sine function.  A non-degenerate pole is thus associated to $e$ such vectors $Q_{1}, \cdots, Q_{e}.$ We denote the collection of these $e$ vectors as $\mathbf{Q}(u_{*},v_{*}).$

For each non-degenerate pole, the Jeffrey-Kirwan residue operator may be conveniently phrased in terms of the associated vectors $\mathbf{Q}(u_{*},v_{*}).$  First, define the positive cone of the vectors $\mathbf{Q}(u_{*},v_{*}),$ to be those vectors that may be expressed as linear combinations of the $Q_{i}$ with positive real coefficients.  We denote this cone as  $\text{Cone}(\mathbf{Q}(u_{*},v_{*})).$  Then, the Jeffrey-Kirwan residue operator is
\begin{align}\label{JK}
\underset{(u,v)=(u_*,v_{*})}{\text{JK-Res}}\left(
\mathbf{Q}(u_*,v_{*}) ,\zeta
\right)
{1 \over Q_{_{1}}(u,v)}\cdots
{1 \over Q_{_{e}}(u,v)}
=
\begin{cases}
\det|(Q_{1}\cdots Q_{e})|^{-1}~,&\zeta\in \text{Cone}(\mathbf{Q}(u_{*},v_{*}))~,\\
0~,&\zeta\notin \text{Cone}(\mathbf{Q}(u_{*},v_{*}))~.
\end{cases}
\end{align}

Thus, to evaluate the general residue formula \eqref{JKform} we must enumerate the non-degenerate poles with $\zeta\in\text{Cone}(\mathbf{Q}(u_{*},v_{*})).$  To determine the full $y$-dependent index of the star quiver \eqref{JKform}, we then sum the resulting contributions from each pole.  We carry this out explicitly in \S \ref{sec:spin}. Meanwhile, in the special case of the Euler characteristic $(y=1)$ it is easy to see that each pole with $\zeta\in\text{Cone}(\mathbf{Q}(u_{*},v_{*}))$ contributes exactly $1/(Mr+1)!$ to $\chi(\mathcal{M}_{m_*,Mr+1}^k).$ Evaluating the Euler characteristic is therefore reduced to the combinatorial enumeration of contributing poles.

We begin our task of tallying these poles by examining the FI parameter. In the case $N=Mr+1$ of interest this takes the form
\begin{align}\label{FI2}
\begin{split}
\zeta&=\left(\underbrace{{1},\cdots, {1}}_{N} ,\, 
\underbrace{ - (r +{1\over M}), \cdots, -(r +{1\over M})}_{m_1-1},\,\underbrace{ -(2r +{2\over M}), \cdots, -(2r +{2\over M})}_{m_2}, \,
\right.\\
&\left.
~~~~~~\cdots,\,
\underbrace{ - (\ell r +{\ell\over M}), \cdots, -(\ell r +{\ell\over M})}_{m_\ell}, \cdots
\right)~.
\end{split}
\end{align}
A contributing pole is defined by choosing $e$ factors in the denominator of \eqref{Z1loop}, and we aim to constrain admissible choices.

First of all, it is easy to see that we can never have a contributing pole where one of the vanishing sine factors is $\sin(u_a-u_b),$ as in that case the associated vector $Q$ is orthogonal to $\zeta$. Next, note that when we write $\zeta$ as a positive linear combination of the $Q$ vectors corresponding to\footnote{Recall that we have decoupled a node that would have had gauge fugacity $v_1^{(1)}$. The factor $\sin(u_a-\zeta^{(1,1)}_i)$ in $Z_{\text{1-loop}}$ corresponds to the  chiral multiplet connected to this decoupled node.} $\sin(u_a - v^{(\ell)}_I -\xi^{(I,\ell)}_i)$ and $\sin(u_a-\xi^{(1,1)}_i)$, the coefficients are less than or equal to 1; otherwise the components of $\zeta$ corresponding to $u_a$ would exceed 1. Examining the coefficient of $v^{(\ell)}_I,$ it then follows that we have to pick at least $\ell r+1$ factors of the form $\sin(u_a - v^{(\ell)}_I -\xi^{(I,\ell)}_i)$ for some $a=1,\cdots,N$ and $i=1,\cdots,\ell k$. 

We now show that $\ell r+1$ is exactly the number of sines we need to pick to correctly account for the coefficient of $v^{(\ell)}_I$. Suppose we have picked $\ell r+1+\delta^{(I,\ell)}$ factors $\sin(u_a-v^{(\ell)}_I -\xi^{(I,\ell)}_i)$ for each  $(I,\ell)$ with $\delta^{(I,\ell)}\ge0$. We argue that $\delta^{(I,\ell)}=0$. We must select the remaining sine factors to be of the form $\sin(u_a - \xi^{(1,1)}_i)$. The remaining number of sine factors to choose is
\begin{align}
\begin{split}
&e - \sum_{(I,\ell)\neq (1,1)}\left( \ell r +1+\delta^{(I,\ell)}\right)
= (r+1)-\sum_{(I,\ell)\neq (1,1)}\delta^{(I,\ell)}~.
\end{split}
\end{align}
Now the sum of the $N$ components corresponding to $u_a$ is
\begin{align}
(r+1)-\sum_{(I,\ell)\neq (1,1)}\delta^{(I,\ell)}
+\sum_{(I,\ell)\neq (1,1)}\left(\ell r +{\ell\over M}\right)
= M r+1 -\sum_{(I,\ell)\neq (1,1)}\delta^{(I,\ell)}~.
\end{align}
On the other hand, the sum of the components corresponding to $u_a$ for the FI parameter $\zeta$ \eqref{FI2} is $N=Mr+1$. It follows that $\delta^{(I,\ell)}=0$. 

To summarize, for each $(I,\ell),$ we have to pick precisely $\ell r+1$ $\sin(u_a-v_I^{(\ell)}- \xi^{(I,\ell)}_i)$ factors. Meanwhile, we have to pick $r+1$ factors of the form $\sin(u_a- \xi^{(1,1)}_i)$ which we may view as being associated to the decoupled node.  This is the same number had the node not been decoupled. We thus see that in the case $N=Mr+1$, the decoupled node is treated on the same footing as the others. This gives the Jeffrey-Kirwan rule:

\paragraph{JK Rule:} For a given $(I,\ell)$, we need to pick exactly $(\ell r+1)$ factors $\sin (u_a-v_I^{(\ell)}-\xi^{(I,\ell)}_i)$'s in the denominator of \eqref{Z1loop} for the pole to make a non-zero contribution to the residue.

\subsection{From Poles to Bipartite Graphs}
\label{sec:polegraph}

We are now ready to enumerate all the poles satisfying the Jeffrey-Kirwan rule. It turns out that the resulting combinatorics problem has a natural interpretation in graph theory. In this section we describe the dictionary between a contributing pole and a bipartite graph.

We begin with the choices of the flavor fugacities $\xi_{i}^{(I,\ell)}$ appearing in the sine factors. Different choices of flavor fugacities will \textit{not} be reflected as different bipartite graphs which we introduce. Given a group of arrows emanating from the node $(I,\ell)$, the choices for the flavor fugacities are easy to enumerate and will be considered a trivial factor.
 
Indeed, the only rule for choosing the  flavor fugacities is that if the $\xi$'s in two sine factors are the same, then the residue for that pole is zero. For example, if we pick both
\begin{align*}
\sin (u_1- v^{(\ell)}_I- \xi^{(I,\ell)}_1)~,~~~~
\sin (u_2- v^{(\ell)}_I- \xi^{(I,\ell)}_1)~,
\end{align*}
it follows that $u_1=u_2$ at this pole. The residue for this pole is then zero due to the factor $\sin^2(u_1-u_2)$ in the numerator of \eqref{Z1loop}. Hence for each group of arrows labels by $(I,\ell)$, we have to pick $\ell r+1$ distinct $\xi$ (with order) out of the $\ell k$ of them. This results in the following factor
\begin{align}\label{countflavor}
\prod_{\ell=1}^M \left[ { \ell k\choose \ell r+1} (\ell r+1)!\right]^{m_\ell}.
\end{align}

Next, we must choose which $u_a$'s are in the selected $\sin(u_a - v_I^{(\ell)} - \xi)$ factors.\footnote{We will not write the superscript for the flavor fugacities from now on as the number of choices have been enumerated in \eqref{countflavor}.} We represent these choices by a bipartite graph $G=(V_L+V_R,E)$ with $V_L$ and $V_R$ being the two disjoint sets of vertices and $E$ being the set of edges. 

The set of bipartite graphs of interest is $\mathcal{G}(\vec{L}_{m_*,r})$ introduced in \S \ref{subsec:graph}. For a bipartite graph in $\mathcal{G}(\vec{L}_{m_*,r})$, it has $v_L=\sum_\ell m_\ell $ vertices in $V_L$ and $v_R= N=Mr+1$ vertices in $V_R$. We will label the vertex in $V_L$ by a pair of integers\footnote{If $m_\ell=0$, then we do not have the corresponding vertex.} $(I,\ell)$ with $I=1,\cdots, m_\ell$ and $\ell=1,\cdots,M$. For $G\in \mathcal{G}(\vec{L}_{m_*,r})$, the $(I,\ell)$-th vertex in $V_L$ is incident to $\ell r+1$ edges. The total number of edges $e$ is therefore
\begin{align}\label{edges}
e= \sum_{\ell=1}^M m_\ell (\ell r+1) = Mr +\sum_{\ell=1}^M m_\ell,
\end{align}
which is again the total rank of the star quiver.

\begin{figure}[h!]
\begin{center}
\includegraphics[width=.65\textwidth]{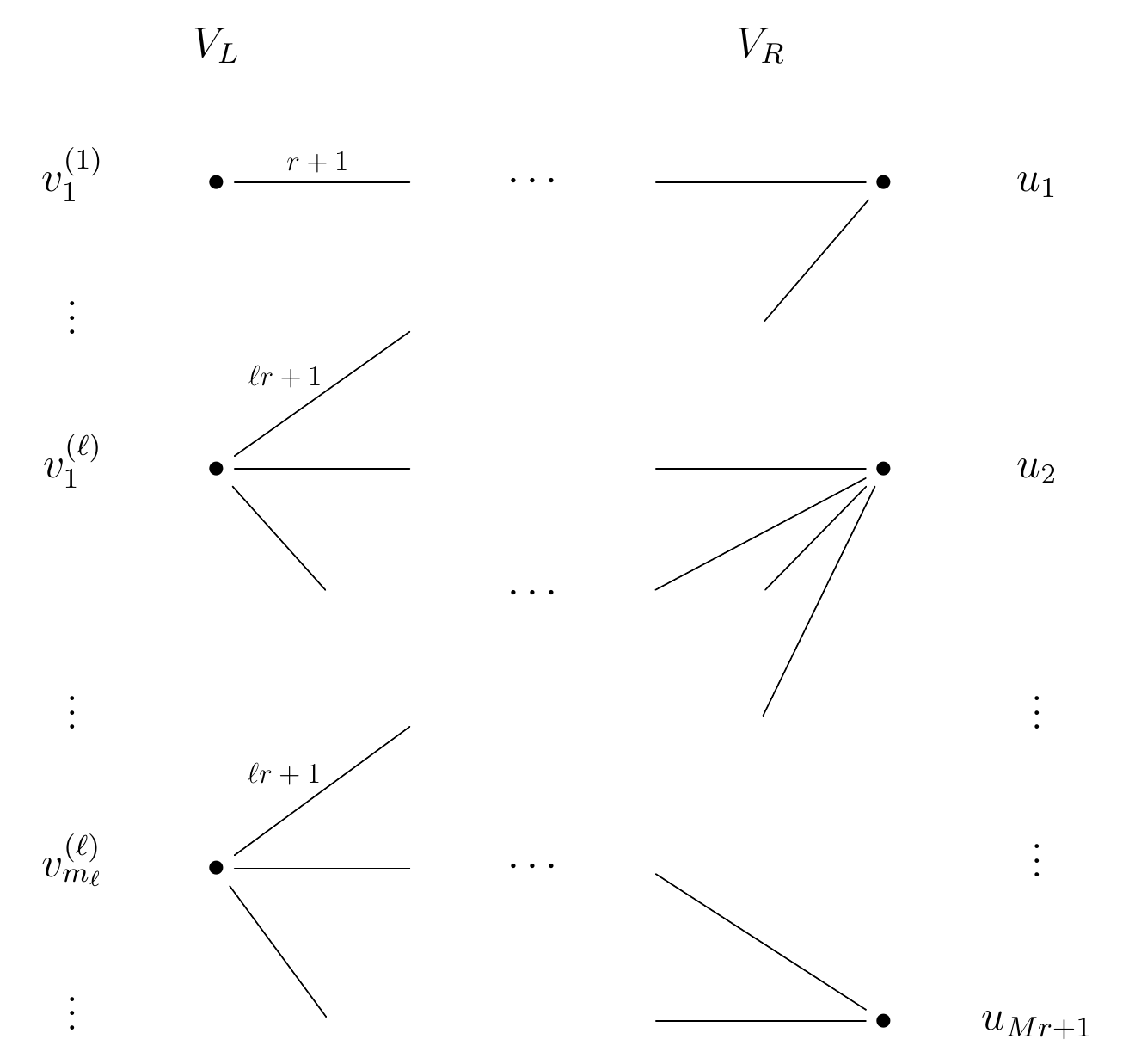}
\end{center}
\caption{Each pole (modulo the flavor fugacity degeneracy \eqref{countflavor}) for the star quiver $\mathcal{M}^k_{m_*,Mr+1}$ satisfying the Jeffrey-Kirwan rule corresponds to a  bipartite graph $G$ in the set $\mathcal{G}(\vec{L}_{m_*,r})$ with $\vec{L}_{m_*,r}$ given in \eqref{Lmj}. Each vertex in $V_L$ corresponds to a gauge fugacity for the abelian node $v_I^{(\ell)}$, with $I=1,\cdots, m_\ell$ ($v_I^{(\ell)}$ does not exist if $m_\ell=0$), while each vertex in $V_R$ corresponds to a gauge fugacity for the non-abelian node $u_a$, $a=1,\cdots, Mr+1$. There are $v_L=\sum_\ell m_\ell$ vertices in $V_L$ and $v_R = Mr +1$ vertices in $V_R$. The $(I,\ell)$-th vertex in $V_L$ is incident to $\ell r+1$ edges. The number of edges for a vertex $u_a$ in $V_R$ is not constrained. The total number of edges is $e= Mr + \sum_\ell m_\ell$. The main quantity of interest $T(\vec{L}_{m_*,r})$ is the number of trees (connected graphs with no cycle) in this set of bipartite graphs $\mathcal{G}(\vec{L}_{m_*,r})$.   }\label{fig:graphproblem2}
\end{figure}

We may now translate every pole for the star quiver associated to the partition $m_*$  into a bipartite graph in $\mathcal{G}(\vec{L}_{m_*,r})$,
\begin{itemize}
\item Each vertex in $V_L$ corresponds to a gauge fugacity $v_I^{(\ell)}$ of the $(I,\ell)$-th abelian node. 
\item Each vertex in $V_R$ corresponds to a gauge fugacity  $u_a$ of the non-abelian node.
\item  An edge between the $(I,\ell)$-th vertex in $V_L$ and the $a$-th vertex in $V_R$ corresponds to a pole satisfying $u_a- v_I^{(\ell)}-\xi=0$.
\end{itemize} 
To specify a non-degenerate pole, we have to pick $e$ sine factors. In the language of bipartite graph, we have to draw precisely $e$ edges. The resulting geometry is indicated in Figure \ref{fig:graphproblem2}.

\begin{figure}[h]
\begin{center}
\includegraphics[width=.35\textwidth]{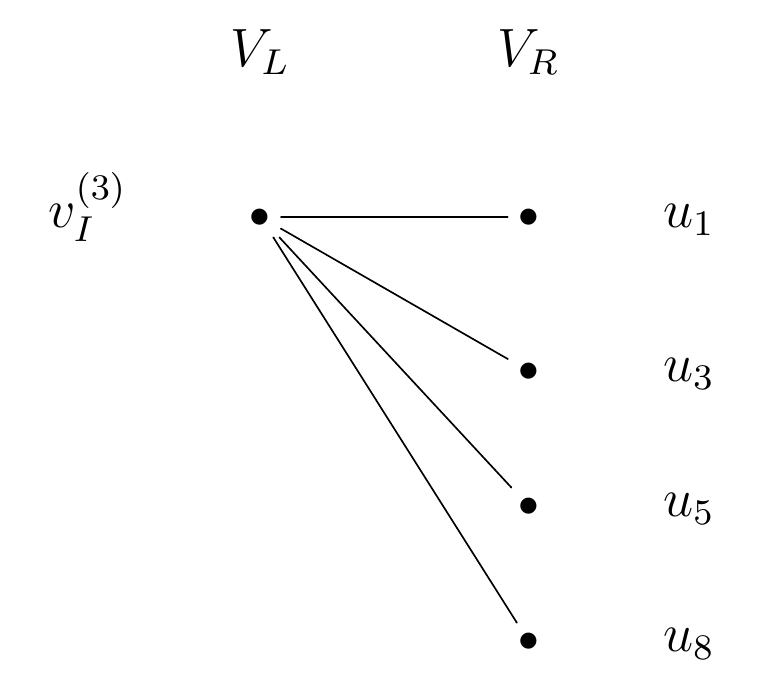}
\end{center}
\caption{ A bipartite subgraph corresponds to the choice of vanishing factors in \eqref{example1}.}\label{fig:example1}
\end{figure}

As an example of the correspondence between poles and graphs, consider for $r=1$ and $\ell=3$, a choice of denominator factors
\begin{align}\label{example1}
\sin(u_1- v^{(3)}_I -\xi)~,~~~\sin(u_2- v^{(3)}_I -\xi)~,~~~\sin(u_5- v^{(3)}_I -\xi)~,~~~\sin(u_8- v^{(3)}_I -\xi)~.
\end{align}
This corresponds to the bipartite subgraph in Figure \ref{fig:example1}. Note that we have $\ell r+1=4$ edges incident to the vertex $v_I^{(3)}$.

\begin{figure}[h!]
\begin{center}
\includegraphics[width=.35\textwidth]{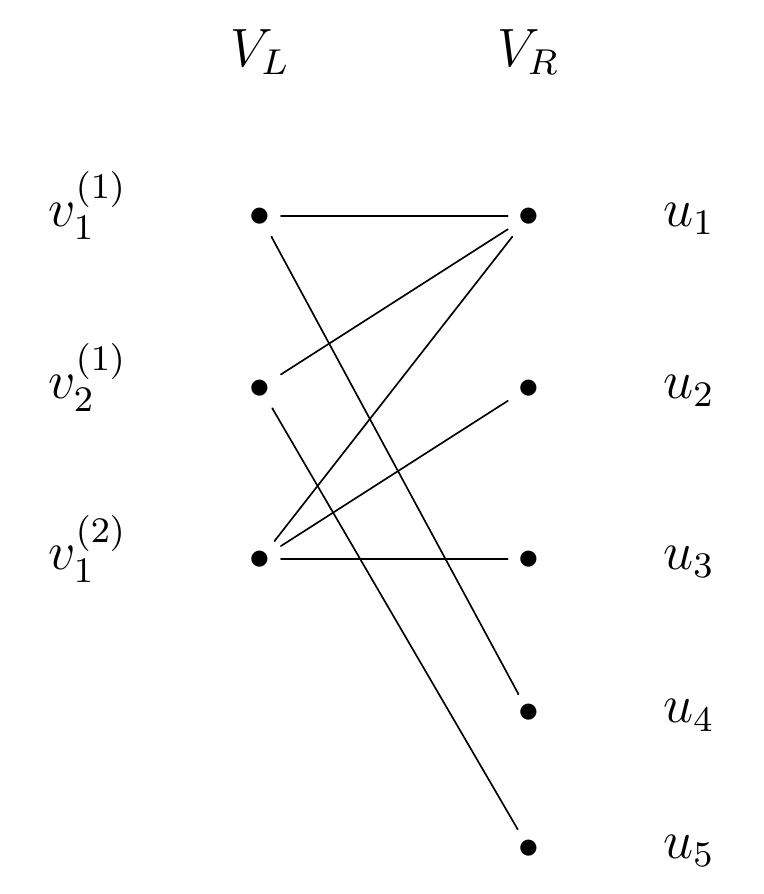}
\end{center}
\caption{ The bipartite graph corresponding to the pole in \eqref{example2}. In this example $M=4$, $r=1$, and the partition is $m_* = (2,1,0,0)$. The number of vertices are $v_L=3$ and $v_R=5$. There are 2 edges incident to the vertices $v_1^{(1)}$ and $v_2^{(1)}$ and 3 edges incident to the vertex $v_1^{(2)}$. }\label{fig:example2}
\end{figure}

As a more advance example, consider  
\begin{align}
r=1,~~ M=4,~~m_*=\left( m_1=2,\,m_2=1,\,m_3=0,\,m_4=0\right).
\end{align}
 The number of vertices $v_I^{(\ell)}$ in $V_L$ and the number of vertices $u_a$ in $V_R$ are
 \begin{align}
 v_L =\sum_\ell m_\ell =3,~~~v_R=Mr+1=5.
 \end{align}
 There are 2 edges incident to the vertices $v_1^{(1)}$ and $v_2^{(1)}$ and 3 edges incident to the vertex $v_1^{(2)}$.   An example of poles satisfying the Jeffrey-Kirwan rule is given by the choice of sine factors\footnote{As usual, $v_1^{(1)}$ is understood to be zero below, though we sometimes write it explicitly to make the notation more uniform.}
\begin{align}\label{example2}
\begin{split}
&\sin(u_1   - v^{(1)}_1-\xi )~,~~~\sin(u_4  -v_1^{(1)}-\xi )~,\\
&\sin(u_1 - v^{(1)}_2 -\xi )~,~~~\sin(u_5 - v^{(1)}_2 -\xi )~,\\
&\sin(u_1 - v^{(2)}_1 -\xi )~,~~~\sin(u_2 - v^{(2)}_1 -\xi )~,~~~\sin(u_3 - v^{(2)}_1 -\xi )~.
\end{split}
\end{align}
This choice corresponds to the bipartite graph in $\mathcal{G}(\vec{L}_{m_*,1})$ in Figure \ref{fig:example2}.

\subsection{From Bipartite Graphs to Poles: The Tree Rule}
\label{sec:graphpole}

So far we have translated every  pole (modulo the degeneracy of choosing the flavor fugacities \eqref{countflavor})  satisfying the Jeffrey-Kirwan rule into a bipartite graph in $\mathcal{G}(\vec{L}_{m_*,r})$.  However not every bipartite graph in $\mathcal{G}(\vec{L}_{m_*,r})$ has an associated contributing pole in the index problem.  As we argue in this section, only those $G\in \mathcal{G}(\vec{L}_{m_*,r})$ that are connected and have no cycles, known as \textit{trees} in graph theory, correspond to poles that exist. This will be referred as the ``Tree Rule".

\begin{figure}[h]
\begin{center}
\subfloat[]{
\includegraphics[width=.3\textwidth]{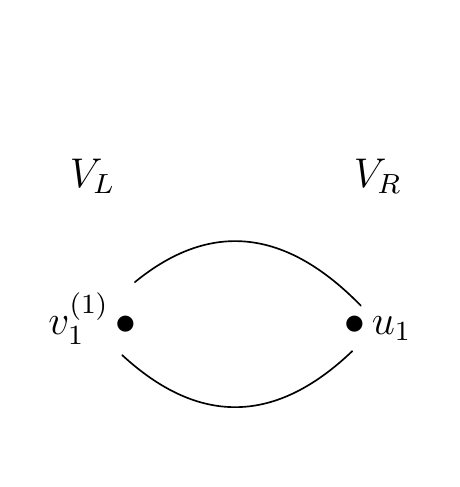}
}
~~~~~~~~~~~~~~~~
\subfloat[]{
\includegraphics[width=.3\textwidth]{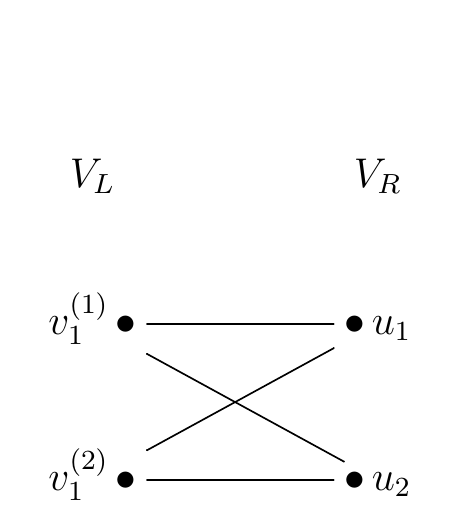}
}
\end{center}
\caption{ Two bipartite graphs corresponding to \eqref{example31} and \eqref{example32}. They do not represent actual poles  because they are not trees (connected graphs with no cycle).}\label{fig:example3}
\end{figure}

The essential logic may be demonstrated in a basic example. For instance, for $r=1$, the candidate pole
\begin{align}\label{example31}
&\sin(u_1  -\xi^{(1,1)}_i )~,~~~\sin(u_1  -\xi^{(1,1)}_j )~,  
\end{align}
for some $i\neq j$ corresponds to the bipartite graph in Figure \ref{fig:example3}(a).  At this hypothetical pole we have $u_1-\xi^{(1,1)}_i=0$ and $u_1-\xi^{(1,1)}_j=0.$  However, this implies a relation among the flavor fugacities and violates our choice of these parameters as generic complex numbers \eqref{rationalrel}.   Thus, there is in fact no pole associated to the choice of vanishing factors \eqref{example31}.  The key observation is that the relation on the flavor fugacities arises from a cycle in the corresponding graph.

As a more advanced example with $r=1,$ consider the hypothetical choice of vanishing factors
\begin{align}\label{example32}
\begin{split}
&\sin(u_1  - v_1^{(1)} - \xi^{(1,1)}_i )~,~~~\sin(u_2  -v_1^{(1)} - \xi^{(1,1)}_j )~,\\
&\sin(u_1  -v_2^{(1)}-\xi^{(2,1)}_k )~,~~~\sin(u_2  -v_2^{(1)}-\xi^{(2,1)}_\ell )~.
\end{split}
\end{align}
At this candidate pole one then has
\begin{align*}
u_1-u_2 = \xi_i^{(1,1)} -\xi^{(1,1)}_j  = \xi^{(2,1)}_k - \xi^{(2,1)}_\ell~,
\end{align*}
which is again in contradiction with our generic choice of the flavor fugacities \eqref{rationalrel}.  Thus there is no such pole.  Again we see that there is a cycle in the associated graph Figure \ref{fig:example3}(b).

The phenomenon illustrated in these examples is general.  Any graph with a cycle does not correspond to an actual pole because the associated choice of vanishing factors implies a false relation on the flavor fugacities. Indeed, it is easy to construct such a relation by summing the arguments of the sine factors associated to the edges in the cycle with alternating signs.

In conclusion a pole exists and contributes to the index if and only if the associated bipartite graph $G\in \mathcal{G}(\vec{L}_{m_{*},r})$ has no cycles. Since the number of edges is $e =v_L+v_R-1$ an elementary result in graph theory (Proposition \ref{prop:tree} in \S \ref{sec:graph}) then shows that the graph is in fact a tree \textit{i.e.} a connected graph with no cycle. We have hence arrived at our Tree Rule:

\paragraph{Tree Rule:} A bipartite graph in $\mathcal{G}(\vec{L}_{m_*,r})$ corresponds to a pole if and only if it is a tree.
\paragraph{}
This completes our graph theoretic enumeration of poles contributing to the residue formula \eqref{JKform}.  To determine the full $y$ dependent index of the star quiver (and hence also the Kronecker quiver) we must sum the resulting contributions from each tree.  We carry out this procedure in \S \ref{sec:spin} for various examples.    

In the case of the Euler characteristic, the results of this section provide more complete information.  Given a star quiver associated to the partition $m_*$ and a non-negative integer $r$, the number of contributing poles is the number of trees in the set $\mathcal{G}(\vec{L}_{m_*,r})$, which we call $T(\vec{L}_{m_*,r})$. Each pole contributes to the Euler characteristic by $1/(Mr+1)!$. Combining these with the choices of flavor fugacities \eqref{countflavor}, we arrive at the formula
\begin{align}\label{stareulerform}
\chi(\mathcal{M}_{m_*,Mr+1}^k) = {1\over (Mr+1)!}  
T(\vec{L}_{m_*,r})  \prod_{\ell=1}^M \left[ { \ell k\choose \ell r+1} (\ell r+1)!\right]^{m_\ell},
\end{align}
with $T(\vec{L}_{m_*,r})$ the number of trees with incidence data as described in \S \ref{subsec:graph}.

\section{Graph Theory}\label{sec:graph}

In this section we will prove the Tree Counting Theorem \ref{theorem} in graph theory language.  Combined with the results of \S \ref{sec:derivation}, this completes the derivation of our main result \eqref{final3} for the Euler characteristic  of Kronecker moduli space.

In \S \ref{sec:general} we introduce necessary basic notions in graph theory.  In \S \ref{sec:division} we define the concept of \emph{division} of a bipartite graph which crucial to our proof.  Finally, in \S \ref{sec:graphproof} we prove the Tree Counting Theorem using divisions.

\subsection{Generalities in Graph Theory}
\label{sec:general}

We begin with a review of various basic notions in graph theory.  We denote the number of vertices and edges of a graph $G$ by $v$ and $e$, respectively. All the graphs of interest are \textit{undirected} graphs (the edges are unorientated).

\begin{definition}\label{def1}
Let $G$ be a graph consisting of vertices and  edges. A \textbf{trail} is a sequence of vertices and edges in $G$, where  each edge's endpoints are the preceding and following vertices in the sequence. A \textbf{path} is a trail where no vertices (and hence edges) are repeated, except possibly the first and the last. A \textbf{cycle} is a path in which the first and the last vertices are the same.
\end{definition}

\begin{definition}
Let $G$ be a graph. $G$ is a \textbf{tree} if it is  connected and has no cycle.
\end{definition}

\begin{proposition}\label{prop:tree}
The following are equivalent for a graph $G$ with $e$ edges and $v$ vertices:
\begin{itemize}
\item $G$ is a tree.
\item $G$ is connected and $e=v-1$.
\item $G$ has no cycle and $e=v-1$.
\end{itemize}
 \end{proposition}
\paragraph{Proof} See, for example, Theorem 3.1.3 of \cite{Gould}.\\

The  graphs of most interest for our purpose are bipartite graphs:
\begin{definition}
A \textbf{bipartite graph} is a graph whose vertices can be divided into two disjoint sets $V_L$ and $V_R$ such that every edge connects a vertex in $V_L$ to $V_R$. $V_L$ and $V_R$ are called the \textbf{partite sets}.
\end{definition}

We denote a bipartite graph by $G=(V_L+V_R,E)$ with $E$ being the set of edges in the graph. We denote the number of vertices in $V_L$ and $V_R$ by $v_L$ and $v_R$, respectively.  As before, the total number of vertices and edges are indicated by $v(=v_L+v_R)$ and $e$, respectively.

\subsection{Division of Bipartite Graphs}
\label{sec:division}

In this section we develop various concepts which are useful preliminaries to the Tree Counting Theorem \ref{theorem}. The main idea is called division defined below.

\begin{definition}[Division]
 Let $G=(V_L+V_R,E)$ be a bipartite graph. A \textbf{division} of $G$ is a disjoint union $E=L\cup R$ of edges such that the following conditions hold.
\begin{itemize}
\item Each vertex in $V_L$ is incident to at most one edge in $L$ ({\color{blue}blue}).
\item Each vertex in $V_R$ is incident to at most one edge in $R$ ({\color{red}red}).
\end{itemize}
Each of $L$ and $R$ will be called a \textbf{compartment} of the division. We will put the edges in $L$ and $R$ in blue and red, respectively. $G$ is said to be \textbf{divisible} if it admits a division.
\end{definition}

\begin{figure}[h]
\begin{center}
\raisebox{-11pt}{
\includegraphics[width=0.25\textwidth]{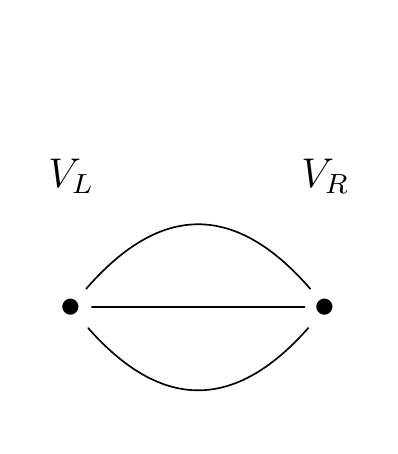}
}
\end{center}
\caption{ A bipartite graph that is not divisible. If it were divisible, then two of the three edges must belong to the same compartment, say, $L$. This leads to a contradiction because the vertex in $V_L$ is incident to two edges in $L$. }\label{fig:nodivision}
\end{figure}

\paragraph{Remark} Not every bipartite graph is divisible. See Figure \ref{fig:nodivision} for a simple example.

\begin{figure}[h!]
\centering
\subfloat[]{
\includegraphics[width=.2\textwidth]{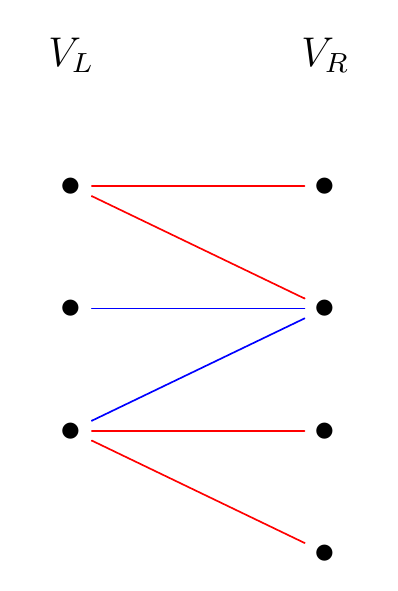}
}~~~~~
\subfloat[]{
\includegraphics[width=.2\textwidth]{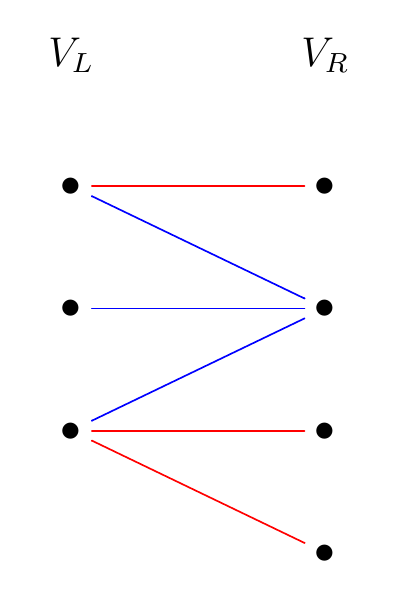}
}
\caption{ (a) An example of a division, where each vertex in $V_L$ is incident to \textit{at most} one edge in $L$ (blue) while each vertex  in $V_R$ is incident to \textit{at most} one edge in $R$ (red). (b) An example of a left maximal division of the same bipartite graph. In a left maximal division every vertex in $V_L$ is incident to \textit{exactly one} edge in the compartment $L$ (blue).
 }
\end{figure}

\paragraph{Remark} Let the numbers of edges in $L$ and $R$ be $e_L$ and $e_R$, respectively. Clearly, $e_L$ ($e_R$ resp.) cannot be bigger than $v_L$ ($v_R$ resp.).

\begin{definition}[Maximal division]
 A division $E=L\cup R$ of $G=(V_L+V_R,E)$ is said to be \textbf{left maximal} if each vertex in $V_L$ is incident to exactly one edge in $L$, \text{i.e.} $e_L=v_L$. Similarly for the \textbf{right maximal} division.
\end{definition}

\begin{proposition} 
Let $G$ be a bipartite tree.  Then there is no division of $G$ that is simultaneously left and right maximal.
\end{proposition}

\paragraph{Proof} Suppose there is a simultaneous left and right maximal division of bipartite tree $G$, then $e_L=v_L$ and $e_R=v_R$. The total number of edges is 
\begin{align}
e=e_L+e_R~,
\end{align}
which is equal to the total number of vertices $v_L+v_R$. By Proposition \ref{prop:tree}, $G$ cannot be a tree, hence a contradiction. $\square$

\begin{theorem} \label{thm:LMD}
 Let $G=(V_L+V_R,E)$ be a tree. Then $G$ has exactly $v_R$ left maximal divisions.
 \end{theorem}

\paragraph{Proof}

We will prove this by explicit constructions of the $ v_R$ left maximal divisions, each of which corresponds to a vertex in $V_R$.

Pick a vertex $v$ in $V_R$. Let the edges incident to $v$ be $\{e^{(1)},e^{(2)},\cdots, e^{(Q)}\}$ for some positive integer $Q$.  For each edge $e^{(i)}$, we will construct a subgraph $T_v^{(i)}$ equipped with a left maximal division by the following algorithm. First, color the edge $e^{(i)}$ in blue, \textit{i.e.} assign the edge $e^{(i)}$ to the compartment $L$. Denote the other endpoint of $e^{(i)}$ by $v_1^{(i)}$. Since $v_1^{(i)}$ is already incident to a blue edge (\textit{i.e.} an edge in $L$), the remaining edges, if any, that are incident to $v_1^{(i)}$ must be in red (\textit{i.e.} belonging to $R$). Let us denote these red edges by $ e'^{(i)}_{1},\cdots, e'^{(i)}_{n}$ and the vertices in $V_R$ they are incident to by $v'^{(i)}_{1}, v'^{(i)}_{2},\cdots, v'^{(i)}_{n}$. Now for each of the vertices $v'^{(i)}_{j}$ in $V_R$, since it is already incident to a red edge $e'^{(i)}_{j}$, the remaining edges, if any, that are incident to $v'^{(i)}_{j}$ must be in blue. We repeat this argument and continue the construction till the point when every terminal vertex is incident to only one edge. In this way, for each edge $e^{(i)}$ incident to $v$, we have constructed a subtree $T_v^{(i)}$ equipped with a left maximal division.

Let us note the following two properties of the subtree $T_v^{(i)}$. 
\begin{itemize}

\item First, the intersection of two different subtrees $T_v^{(i)}$  consists only of the original vertex $v$,
\begin{align}
T_v^{(i)} \cap T_v^{(j)} = \{v\}~,~~~~i\neq j~.
\end{align}
Suppose not, then there is another vertex $v'\in T_v^{(i)}\cap T_v^{(j)}$ and $v'\neq v$. We can then form a  cycle passing through $v$ and $v'$ in the union $T_v^{(i)}\cup T_v^{(j)}$, hence contradicting the fact that $G$ is has no cycle.

\item Second, the union of $T_v^{(i)}$ covers the whole graph $G$,
\begin{align}
G= \bigcup_i T_v^{(i)}~,~~~\forall \,v\in V_R~.
\end{align}
This follows from the fact that $G$ is assumed to be connected, so there is a path between $v$ and any other point. This path must belong to one of the subtrees $T^{(i)}_v$ by construction.
\end{itemize}

\begin{figure}
\begin{center}
\includegraphics[width=1\textwidth]{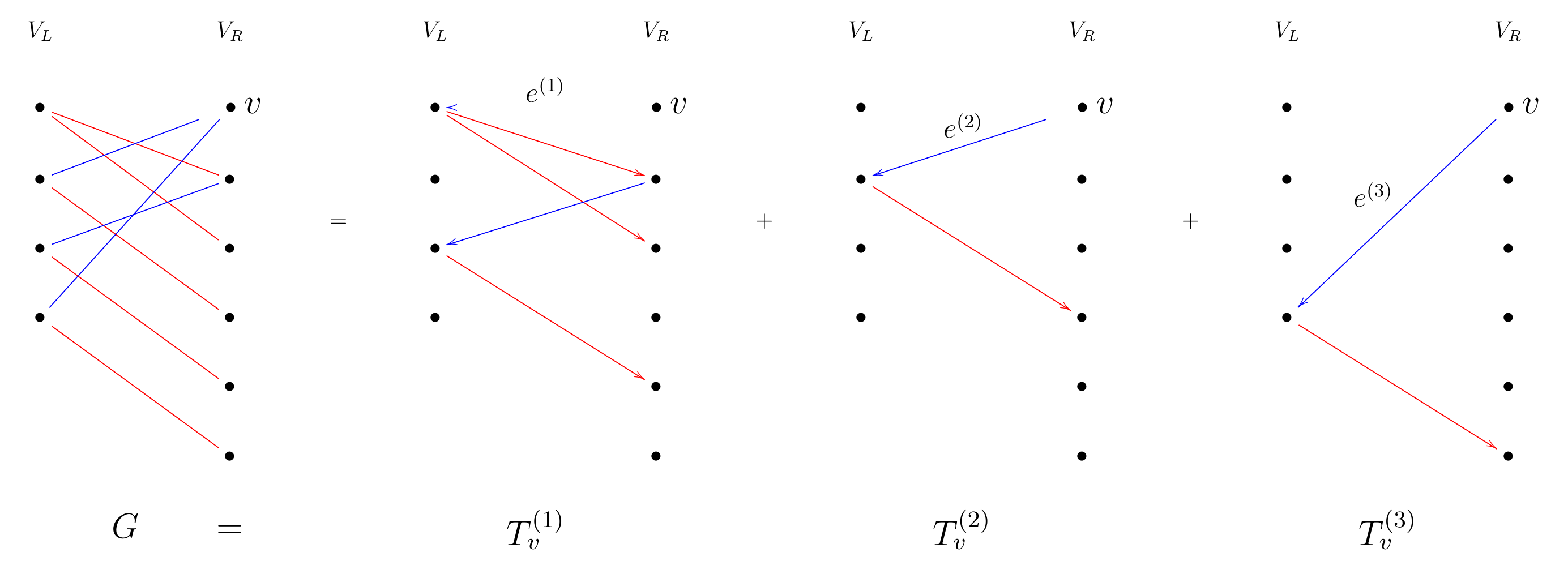}
\end{center}
\caption{Construction of a left maximal division associated to the vertex $v$ of a tree $G$. The graph $G$ is further decomposed into the union of 3 subtrees $T^{(i)}_v$, each of which is associated to an edge $e^{(i)}$ incident to the original vertex $v$. $v$ is the only vertex in $V_R$ that is incident to only blue lines. In a left maximal division, each vertex in $V_L$ is incident to \textit{exactly one} blue line, while each vertex in $V_R$ is incident to \textit{at most} one red line.  In addition to the above left maximal division, there are in total 6 left maximal divisions of $G$, each of which is associated to a vertex in $V_R$. The orientations on the edges indicate the order of the steps in the algorithm.  The graph itself is still undirected.}\label{fig:tree1}
\end{figure}

\begin{figure}
\begin{center}
\includegraphics[width=.55\textwidth]{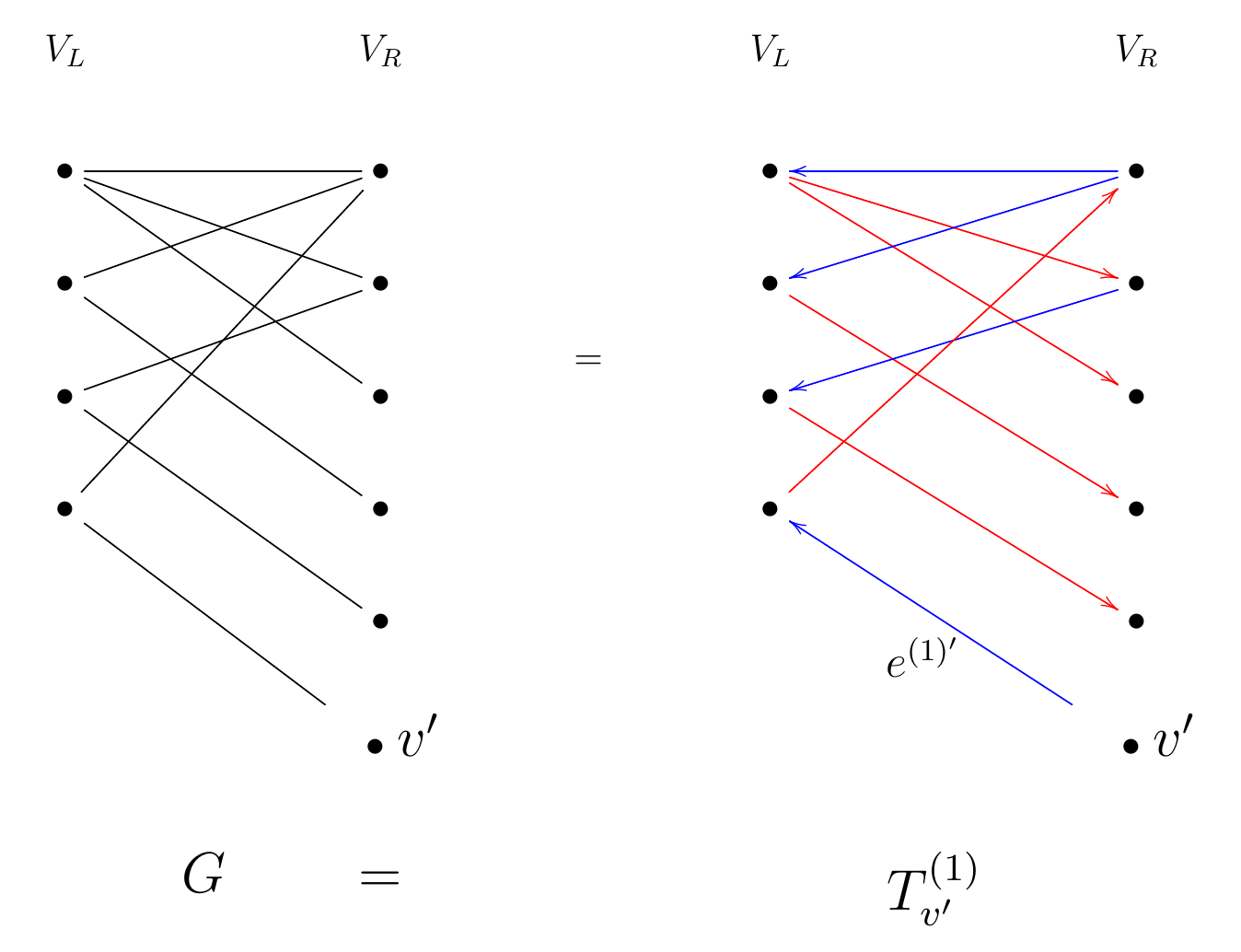}
\end{center}
\caption{Construction of a left maximal division associated to the vertex $v'$ of a tree $G$. In this case the decomposition of $G$ is trivial, $G=T^{(1)}_{v'}$. In a left maximal division, each vertex in $V_L$ is incident to \textit{exactly one} blue line, while each vertex in $V_R$ is incident to \textit{at most} one red line. The orientations on the edges indicate the order of the steps in the algorithm.  The graph itself is still undirected.}\label{fig:tree2}
\end{figure}

Combining the above two properties, we see that for each vertex $v\in V_R$, we can construct a left maximal division of $G$ inherited from that of $T^{(i)}_v$. We will denote this left maximal division of $G$ by $E=L_v\cup R_v$. Note that in $E=L_v\cup R_v$, $v$ is the unique vertex in $V_R$ that is connected  to only blue lines, while each of the other $v_R-1$ vertices is incident to exactly one red line. We illustrate this algorithm for two different vertices in the same bipartite graph in Figure \ref{fig:tree1} and \ref{fig:tree2}.

Next, we want to show that there are no other left maximal divisions of $G$ than $E=L_v\cup R_v$. Let us note that the number of blue lines $e_L$ in a left maximal division is the number of vertices in $V_L$, $e_L=v_L$ by definition. By Proposition \ref{prop:tree}, the number of edges $e$ in a tree is $v_L+v_R-1$. Hence the number of red lines $e_R$ in a left maximal division of a tree is
\begin{align}
e_R = e- e_L = (v_L+v_R-1)  - v_L = v_R-1~.
\end{align}
Since every vertex in $V_R$ is allowed to incident to at most one red line, it follows that among the $v_R$ vertices in $V_R$, there is exactly one of them, say, $v$, that is incident to only blue lines. The colors of the remaining edges are thus determined by the algorithm above to be those of $E=L_v\cup R_v$, and hence there are no  left maximal divisions other than $E=L_v\cup R_v$ for $v\in V_R$. $\square$\\
~

In summary, we have shown that the $v_R$ left maximal divisions $E=L_v\cup R_v$ constructed above are the only left maximal divisions for a tree $G$. In the division $E=L_v\cup R_v$,  $v$ is the unique vertex in $V_R$ that is connected  to only blue lines,  while each of the other $v_R-1$ vertices is incident to exactly one red line. The tree $G$ can be decomposed into unions of smaller trees, $G= \bigcup_i T_v^{(i)}$. Each subtree $T_v^{(i)}$ grows from an edge $e^{(i)}$ incident to $v$. The intersection of two different trees $T_v^{(i)}$ is always the original vertex, 
$T_v^{(i)} \cap T_v^{(j)} = \{v\}$ for $i\neq j$.

\begin{figure}[h]
\begin{center}
\includegraphics[width=.7\textwidth]{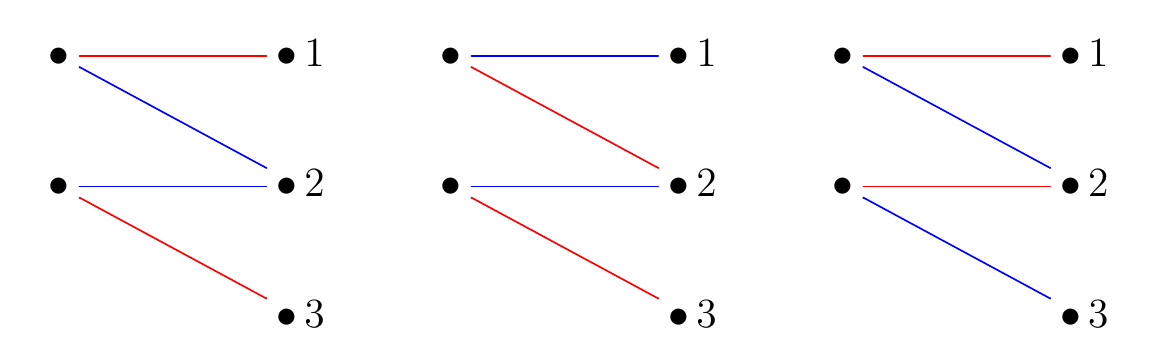}
\end{center}
\caption{The 3 left maximal divisions of an tree bipartite graph in $\mathcal{G}(\vec{L})$, where $\vec{L}=(1,1)\,( =\vec{L}_{\left(m_1=2,\,m_2=0\right),1})$. A bipartite graph in $\mathcal{G}(\vec{L})$ has $v_L=2$ vertices in $V_L$ and $v_R=3$ vertices in $V_R$, with a total number of $e=4$ edges. In a division, each vertex in $V_L$ is incident to at most one edge in $L$ (blue) and each vertex  in $V_R$ is incident to at most one edge in $R$ (red). In the case of left maximal division, each vertex in $V_L$ is incident to exactly one blue edge in $L$.}\label{fig:3maximal}
\end{figure}

\subsection{Proof of the Tree Counting Theorem}
\label{sec:graphproof}

In this section we prove the Tree Counting Theorem \eqref{mainthm}.   From the proof of Theorem \ref{thm:LMD}, we have learned how to construct left maximal divisions for a given bipartite tree  graph $G=(V_L+V_R,E)$. We now reverse the logic to prove the Tree Counting Theorem: we enumerate the total number of left maximal divisions of trees in $\mathcal{G}(\vec{L})$ and then divide by the degeneracy $v_R$ to obtain the number of (uncolored) bipartite tree graphs.

Let us recall the incidence data defining the set of bipartite graphs $\mathcal{G}(\vec{L})$ from a $v_{L}$ component vector $\vec{L}$ with non-negative integral components.
\begin{itemize}
\item There are $v_L$ vertices in $V_L$.
\item There are $v_R=1+\sum_{i=1}^{v_L}L_i$ vertices in $V_R$.
\item There are $L_i+1$ edges incident to the $i$-th vertex in $V_L$.
\end{itemize}
For a tree $G$ in $\mathcal{G}(\vec{L})$, the number of edges $e_L$ and $e_R$ in the compartment $L$ (blue) and $R$ (red) in a left maximal division $E=L_v\cup R_v$ of $G$ is
\begin{align}
e_L = v_L~,~~~~e_R= \sum_{i=1}^{v_L} L_i~.
\end{align}
The $i$-th vertex in $V_L$ is incident to exactly one blue line (being left maximal) and $L_i$ red lines.

\subsubsection*{Strategy of the Proof}

We now enumerate left maximal divisions in $\mathcal{G}(\vec{L})$ that are trees and thereby obtain the number of (uncolored) trees. The procedure is outlined below:
\begin{itemize}
\item \textbf{Step 1}~~Draw  $L_i$ red lines from the $i$-th vertex in $V_L$. Connect the $\sum_i L_i$ red lines to vertices in $V_R$, with each one of them incident to \textit{at most} one red line. We refer to the resulting graph as the \emph{red graph}.
\item \textbf{Step 2}~~Draw exactly one blue line from each vertex in $V_L$. Connect the $v_L$ blue lines to vertices in $V_R$ such that the whole graph is connected. Up to this step we have constructed a left maximal division of a tree in $\mathcal{G}(\vec{L})$.
\item \textbf{Step 3}~~Consider all the left maximal divisions constructed above. Identify those left maximal divisions that correspond to the same (uncolored) tree. Divide the number of connected left maximal divisions by this degeneracy to get $T(\vec{L})$.
\end{itemize}

Note that by Proposition \ref{prop:tree}, in the case of $e=v_L+v_R-1$, a graph is connected if and only if it has no cycles. Therefore, in order to enumerate the trees, it suffices to ensure the connectivity of the graph. Also note that we never generate cycles at Step 1, so it suffices to impose the connectivity condition at Step 2.

Given a red graph $G_R$ constructed in Step 1, let the number of \textit{connected} left maximal divisions in $\mathcal{G}(\vec{L})$ that contain $G_R$ be $B(\vec{L})$. Clearly $B(\vec{L})$ does not depend on the choice of the red graph $G_R$ because every red graph has the same topology. We have the following lemma:
\begin{lemma}\label{lemma:ratio}
Given a red graph $G_R$ constructed in Step 1, the number of connected left maximal divisions in $\mathcal{G}(\vec{L})$ that contain $G_R$ is
\begin{align}
B(\vec{L})= v_R^{v_L-1}~,
\end{align}
where $v_R= 1+\sum_{i=1}^{v_L} L_i$. In particular, $B(\vec{L})$ does not depend on the choice of the red graph $G_R$.
\end{lemma}
Before proving this lemma, let us note that we are now ready to prove Theorem \ref{theorem}.

\subsubsection*{Proof of the Tree Counting Theorem }

 Starting from \text{Step 1}, there are $e_R=\sum_i L_i$ red lines whereas there are $v_R=1+\sum_iL_i$ vertices in $V_R$, each of which is allowed to be incident to \textit{at most} one red line. It follows that exactly one vertex in $V_R$ has no red line incident to it. Hence, the number of red graphs constructed in \text{Step 1} is
\begin{align}
v_R {(v_R-1)!\over\prod_{i=1}^{v_L}L_i!}~,
\end{align}
where $v_R$ comes from choosing which vertex in $V_R$ is left out and the multinomial coefficient ${(v_R-1)!\over\prod_{i=1}^{v_L}L_i!}$ comes from permuting the $v_R-1$ red lines.

Next for Step 2, Lemma \ref{lemma:ratio} instructs us to multiply the counting by a factor of $B(\vec{L})=
v_R^{v_L-1}.$

 Finally for Step 3 we divide the counting by the degeneracy of left maximal divisions for a given tree $G\in \mathcal{G}(\vec{L})$, which is simply $v_R$ by Theorem \ref{thm:LMD}. 

Combining all these factors together, we have
\begin{align}
T(\vec{L}) =\underbrace{ {\color{red} v_R {(v_R-1)!\over\prod_{i=1}^{v_L}L_i!}}}_{\text{Step 1}}\times
\underbrace{ {\color{blue} v_R^{v_L-1}}  }_{\text{Step 2}}\times\underbrace{{1\over v_R} }_{\text{Step 3}}~,
\end{align}
where the red (blue) factors are the number of ways to draw the red (blue) lines. Hence we have proved the Tree Counting Theorem \ref{theorem}. $\square$

Thus, to complete the argument, it remains to prove Lemma \ref{lemma:ratio}.

\subsubsection*{Proof of Lemma \ref{lemma:ratio}}

\begin{figure}[h!]
\centering
\subfloat[]{
\includegraphics[width=.9\textwidth]{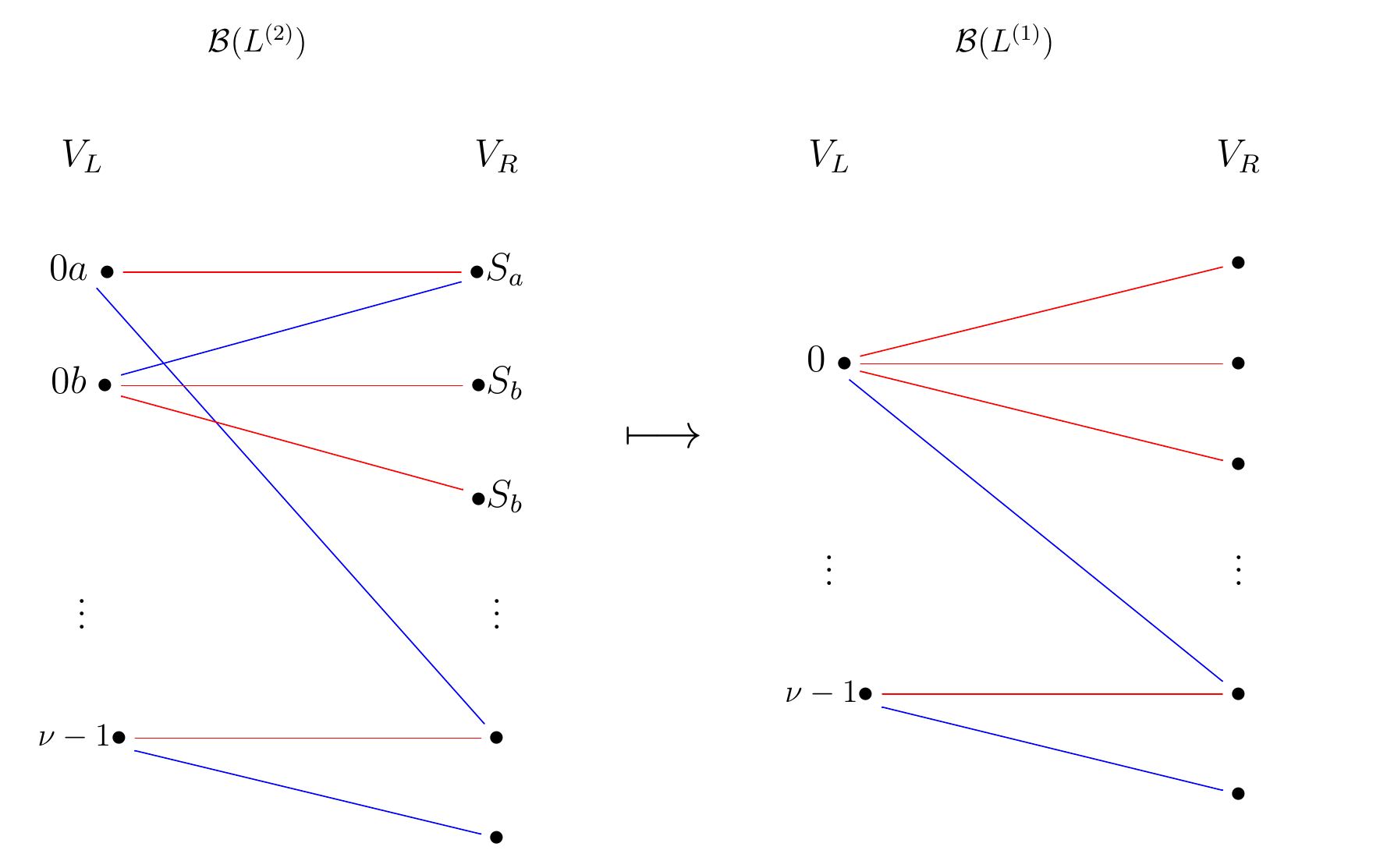}
}\\
\subfloat[]{
\includegraphics[width=.9\textwidth]{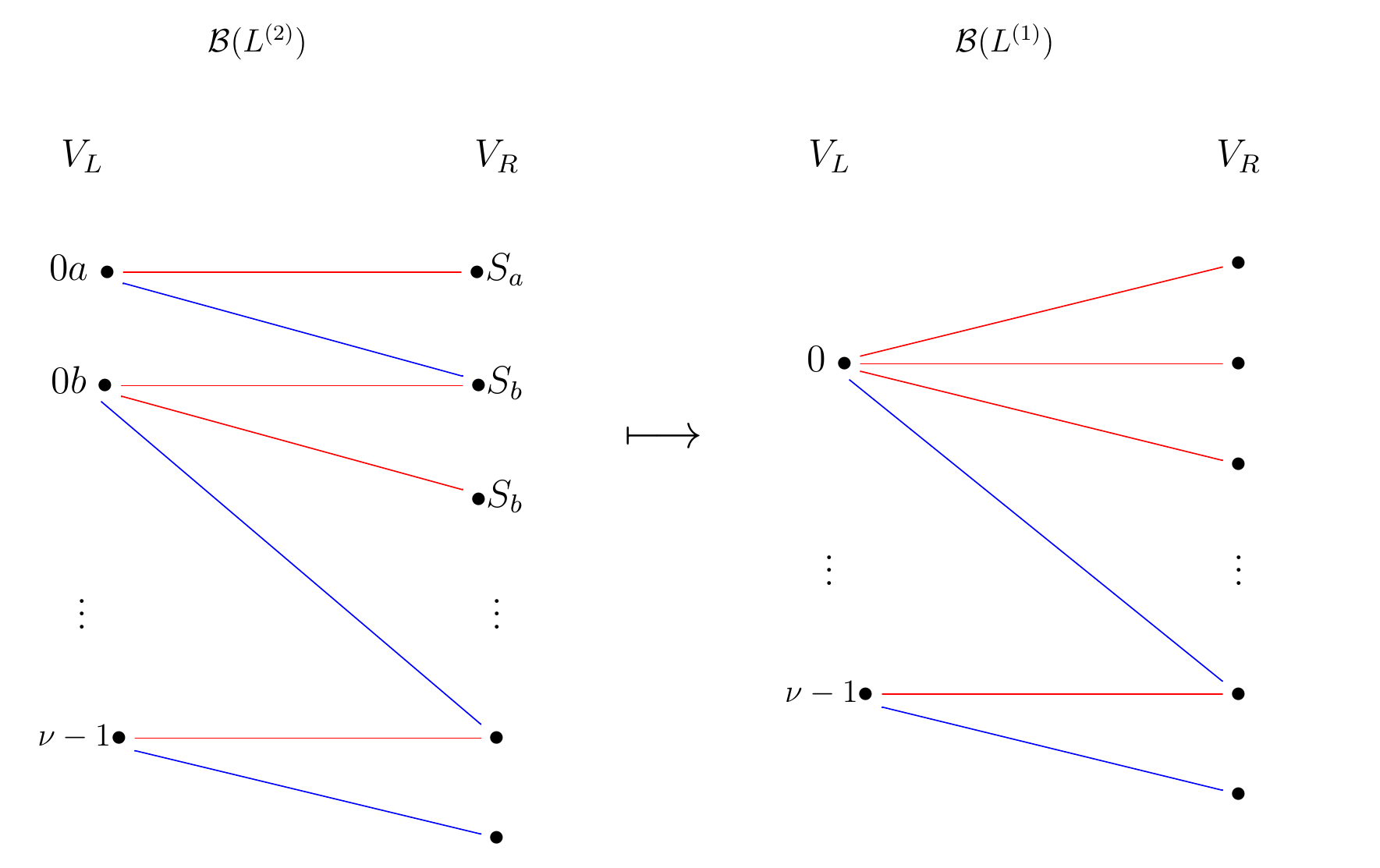}
}
\caption{An illustration of the merging map $\mu:\mathcal{B}(L^{(2)})\rightarrow \mathcal{B}(L^{(1)})$. In this example $L_{0a}=1$ and $L_{0b}=2$. $S_a$ only contains the first vertex in $V_R$ while $S_b$ consists of the second and third vertices in $V_R$. (a) The case $N_{0a}\notin S_b$.  (b) The case $N_{0a}\in S_b$.   }
\label{bigfig}
\end{figure}

It suffices to show the following claim. Let $\nu$ be a positive integer. Let $L_{0a}, L_{0b},L_1,L_2,$ $\cdots, L_{\nu-1}$ be some non-negative integers. Let $\mathcal{I}$ be the index set  $\mathcal{I}=\{ 0a,0b,1,2,\cdots, \nu-1\}$. Define
\begin{align}
&\vec{L}^{(1)} =\underbrace{ (L_{0a}+L_{0b}, \, L_1,\cdots, L_{\nu-1})}_{\nu}~,\\
&\vec{L}^{(2)} =\underbrace{ (L_{0a}, \, L_{0b},\, L_1,\cdots, L_{\nu-1})}_{\nu+1}~.
\end{align}
We will denote the sets of vertices on the left for graphs in $\mathcal{G}(\vec{L}^{(1)})$ and $\mathcal{G}(\vec{L}^{(2)})$ by $V_L^{(1)}$ and $V_L^{(2)}$, respectively. The number of vertices in $V_L^{(1)}$ and $V_L^{(2)}$ are respectively
\begin{align}
v_L^{(1)}=\nu~,~~~v_L^{(2)}=\nu+1~.
\end{align}
The number of vertices on the right in $\mathcal{G}(\vec{L}^{(1)})$ and $\mathcal{G}(\vec{L}^{(2)})$ are both 
\begin{align}
v_R =1+\sum_{i\in \mathcal{I}}L_i~.
\end{align}
We will therefore denote the sets of vertices on the right in $\mathcal{G}(\vec{L}^{(1)})$ and $\mathcal{G}(\vec{L}^{(2)})$ by the same symbol $V_R$ without  superscripts. See Figure \ref{bigfig} for examples of tree in $\mathcal{G}(\vec{L}_{1})$ and $\mathcal{G}(\vec{L}_{2}).$

\paragraph{Claim}
\begin{align}
{B(\vec{L}^{(2)}) = v_R \, B(\vec{L}^{(1)}) }~.
\end{align}

Observe that this claim suffices to prove the lemma.  Indeed for any $\vec{L}$ we have  $B(\vec{L})=1$ for $v_L=1$.\footnote{For $v_L=1$, the vector has only one component, $\vec{L} = (L_1)$. We can choose a red graph in Step 1 to be that each of the first $L_1$ vertices in $V_R$ is incident to a red line while the last vertex is left out. In order not to have a cycle, the blue line has to be incident to the last vertex in $V_R$, and hence $B(\vec{L})=1$.}  Thus, Lemma \ref{lemma:ratio} follows immediately from the above claim by induction on $v_L$.

\paragraph{Proof of the Claim} 
Let us label the vertices  in $V_R$ by $1,2,\cdots, v_R=1+\sum_{i\in \mathcal{I}} L_i$,
\begin{align}
V_R= \{1,2,\cdots,v_R\}~.
\end{align}
 For reasons that will become clear momentarily, we will label the vertices in $V_L^{(1)}$ and $V_L^{(2)}$ as
\begin{align}
&V_L^{(1)} = \{ 0, 1,2,\cdots, \nu-1\}~,\\
&V_L^{(2)}  =\{ 0a,0b,1,2,\cdots, \nu-1\}~.
\end{align}

Since $B(\vec{L})$ does not depend on the choice of the red graph, we will pick a convenient red graph $G_R^{(1)}$ in $\mathcal{G}(\vec{L}^{(1)})$  such that the first $L_{0a}+L_{0b}$ vertices in $V_R$ are incident to the vertex 0 in $V_L^{(1)}$, the next $L_1$ vertices in $V_R$ are incident to the  vertex 1 in $V_L^{(1)}$ and so on. Similarly, we will pick a convenient red graph $G_R^{(2)}$ in $\mathcal{G}(\vec{L}^{(2)})$  such that the first $L_{0a}$ vertices in $V_R$ are incident to the vertex $0a$ in $V_L^{(2)}$, the next $L_{0b}$ vertices in $V_R$ are incident to the  vertex $0b$ in $V_L^{(2)}$ and so on. The last vertex in $V_R$ is left out in both $G_R^{(1)}$ and $G_R^{(2)}$. Let $S_a\subseteq V_R$ be the set of the first $L_{0a}$ vertices in $V_R$ and $S_b\subseteq V_R$ be the next $L_{0b}$ vertices in $V_R$.

We will denote the set of \textit{connected} (and therefore tree in this case) left maximal divisions in $\mathcal{G}(\vec{L}^{(I)})$ that contain $G_R^{(I)}$ as $\mathcal{B}(\vec{L}^{(I)})$, with $I=1,2$. The number of graphs in $\mathcal{B}(\vec{L}^{(I)})$ is by definition $B(\vec{L}^{(I)})$. We will label a graph $b^{(1)}$ in $\mathcal{B}(\vec{L}^{(1)})$ by a tuple of $\nu$ positive integers as
\begin{align}
b^{(1)}= (n_{0},n_1,\cdots,n_{\nu-1}) \in \mathcal{B}(\vec{L}^{(1)})~,
\end{align}
meaning the  vertex $i$ in $V_L^{(1)}$ is incident to the  vertex $n_i$ in $V_R$ by a blue line. Together with the fact that $b^{(1)}$ contains the red graph $G_R^{(1)}$, this tuple of integers uniquely determines the left maximal division $b^{(1)}$.  Similarly we label a graph $b^{(2)}$ in $\mathcal{B}(\vec{L}^{(2)})$ by a tuple of $\nu+1$ positive integers as
\begin{align}
b^{(2)}= (N_{0a},\,N_{0b},\, N_1,\cdots,N_{\nu-1}) \in \mathcal{B}(\vec{L}^{(2)})~.
\end{align}

Note that  not every choice of integers correspond to a connected graph, and therefore does not belong to $\mathcal{B}(\vec{L}^{(I)})$. For example,  we have
\begin{align}\label{requirement1}
N_{0a}\notin S_a~,~~~\text{if}~~b^{(2)}\in\mathcal{B}(\vec{L}^{(2)})~.
\end{align}
And similarly $N_{0b}\notin S_b$ if $b^{(2)}\in\mathcal{B}(\vec{L}^{(2)})$. Otherwise we can form a cycle using the corresponding blue line and a red line in $G_R^{(2)}$. Also, graphs with $N_{0a}\in S_b,$ and  $N_{0b} \in S_a$ have a cycle.  We therefore have
 \begin{align}\label{requirement2}
 N_{0a}\notin S_b ~~\text{or}~~N_{0b}\notin S_a,~~~\text{if}~~b^{(2)}\in\mathcal{B}(\vec{L}^{(2)}).
 \end{align}
For the same reason we have
\begin{align}\label{requirement3}
n_0\notin S_a\cup S_b,~~~\text{if}~~b^{(1)}\in\mathcal{B}(\vec{L}^{(1)}).
\end{align}
These requirements will be important in the following.

Now define the \textit{merging map} $\mu:\,\mathcal{B}(\vec{L}^{(2)})\rightarrow \mathcal{B}(\vec{L}^{(1)})$ as
\begin{align}
\begin{split}
& \mu:~(N_{0a},\,N_{0b},\,N_1,\cdots,N_{\nu-1}) \mapsto
\begin{cases}
 (n_0= N_{0a},\, n_1=N_1, \,\cdots , \, n_{\nu-1}=N_{\nu-1} ),~~~~\text{if}~~N_{0a}\notin S_b,\\
  (n_0= N_{0b},\,n_1=N_1,\,\cdots , \,n_{\nu-1}=N_{\nu-1} ),~~~~\text{if}~~ N_{0a}\in S_b.
  \end{cases}
\end{split}
\end{align}
See Figure \ref{bigfig} for examples of the merging map. 

We need to make sure the merging map is consistent with \eqref{requirement1}, \eqref{requirement2}, and \eqref{requirement3}. First note that $N_{0a}$ is never in $S_a$ for $b^{(2)}\in \mathcal{B}(\vec{L}^{(2)})$ by \eqref{requirement1}, so we do not have to worry about hitting a point with $n_0=N_{0a}\in S_a$, which does not belong $\mathcal{B}(\vec{L}^{(1)})$ by \eqref{requirement3}. Secondly, in the case when $N_{0a}\in S_b$, from \eqref{requirement2} it follows that $N_{0b}\notin S_a$. From \eqref{requirement1} we know $N_{0b}\notin S_b$. Hence $N_{0b}\notin S_a\cup S_b$ if $N_{0a}\in S_b$. It follows that the image graph $(n_0= N_{0b},\,n_1=N_1,\,\cdots , \,n_{\nu-1}=N_{\nu-1} )$ lies in $\mathcal{B}(\vec{L}^{(1)})$.

It is clear that the merging map $\mu$ is surjective.  We now argue that $\mu$ is $v_R$-to-1. This then proves the claim.

Thus, fix a graph $b^{(1)}= (n_0,n_1,\cdots,n_{\nu-1})$, we wish to show that its preimage consists of exactly $v_R$ graphs. Let $b^{(2)}= (N_{0a},\, N_{0b},\,N_1,\cdots,N_{\nu-1}) \in \mathcal{B}(\vec{L}^{(2)})$ such that $\mu (b^{(2)})=b^{(1)}$. 

If $N_{0a}\notin S_b$, then $N_{0a}$ is fixed to be $n_0$ for $\mu(b^{(2)})=b^{(1)}$, while we can freely choose $N_{0b} \in V_R- S_b$. There are $v_R- L_{0b}$ graphs in the preimage of this kind. On the other hand, if $N_{0a}\in S_b$, then $N_{0b}$ is fixed to be $n_0$ for $\mu(b^{(2)})=b^{(1)}$, while we can freely choose $N_{0a}$ to be any point in $S_b$. There are $L_{0b}$ graphs in the preimage of this kind. Hence, given any graph $b^{(1)}\in \mathcal{B}(\vec{L}^{(1)})$, we have shown that there are $v_R$ graphs in the preimage $\mu^{-1}(b^{(1)})$ $.~\square$

\clearpage

\section{Computations of the Refined Index}
\label{sec:spin}

Thus far in this work we have primarily been concerned with the degeneracy $\Omega(M,N,k).$  In this section, we extend our analysis to the full $y$-dependent index $\Omega(M,N,k,y)$ whose definition we recall here
\begin{equation}
\Omega(M,N,k,y) \equiv \sum_{p=0}^d y^{2p-d}\, h^{p,p}(\mathcal{M}^k_{M,N})~, \label{spinlast}
\end{equation}
where $d$ is the complex dimension of the moduli space
\begin{equation}
d=kMN-M^{2}-N^{2}+1~.
\end{equation}

The $y$-dependent index contains much more information than the simple degeneracy $\Omega(M,N,k).$  As a consequence, it is more challenging to determine.  Nevertheless as we will see, the residue technique described in detail in \S \ref{sec:derivation} can 
be extended to a useful algorithm for computing the full index $\Omega(M,N,k,y)$.  We present explicit formulas for small $M$ and $N$ in equation \eqref{spinanswers}.

To begin the analysis, we first note that the MPS degeneration formula can be generalized to include the full $y$ dependence of the index \cite{Manschot:2010qz}. In the case of the Kronecker quiver with dimension vector $(M,N)$, if we apply the MPS degeneration formula on the node $M$, we have
\begin{align}\label{MPSy}
\Omega(M,N,k,y)=y^{-M(M-1)- d} \sum_{m_*\vdash M } \left[\prod_{\ell=1}^M{1\over m_\ell!} \left( (-1)^{\ell-1}\over \ell[\ell]_{y^2}\right)^{m_\ell} \right]y^{d_{m_*}}\Omega ( \mathcal{M}_{m_*,N}^k,y)~,
\end{align}
where
\begin{align}
[n]_q = {1-q^n\over 1-q}~,
\end{align}
and $d_{m_*}$ is the complex dimension of the moduli space for  the star quiver associated to the partition $m_*\vdash M$,
\begin{align}
d_{m_*} = kMN -\sum_{\ell=1}^Mm_\ell- N^2  +1~. 
\end{align}
We also recall that $z$ is defined by $y=e^{iz}.$ 

Thus, to compute the refined index $\Omega(M,N,k,y),$ it suffices to determine the refined indices of star quivers $\Omega ( \mathcal{M}_{m_*,N}^k,y).$  This can be carried out with the residue formula of \S \ref{sec:derivation}.   As described there, contributing poles in the residue formula correspond to bipartite trees.  Each pole comes with an associated $y$-dependent contribution which we sum to determine the refined index.

In the remainder of this section, we apply the formula to the Kronecker quiver with dimension vector $(2,2r+1)$ with $k$ arrows.  

\subsection{Example: $(2,1)$}\label{sec:2,1}

\begin{figure}[here!]
  \centering
\subfloat{
\xy  0;<1pt,0pt>:<0pt,-1pt>::
(-300,0) *+{2}*\cir<10pt>{} ="1",
(-240,0) *+{N}*\cir<10pt>{} ="2",
(-270, -10) *+{k} ="b",
\ar @{->} "1"; "2"
\endxy}
~~=~~
-${y\over2(1+y^{2})}$~\subfloat{
\xy  0;<1pt,0pt>:<0pt,-1pt>::
(-300,0) *+{1}*\cir<10pt>{} ="1",
(-240,0) *+{N}*\cir<10pt>{} ="2",
(-270, -10) *+{2k} ="b",
\ar @{->} "1"; "2"
\endxy}
~~+~~
${1\over2}$~\subfloat{
\xy  0;<1pt,0pt>:<0pt,-1pt>::
(-300,0) *+{1}*\cir<10pt>{} ="1",
(-300,-17.5) *+{v} ="4",
(-240,-17.5) *+{u} ="5",
(-240,0) *+{N}*\cir<10pt>{} ="2",
(-270,45) *+{1}*\cir<10pt>{} ="3",
(-270, -10) *+{k} ="b",
(-250, 30) *+{k} ="a",
\ar @{->} "1"; "2"
\ar @{->} "3"; "2"
\endxy}\\

~~~~~~~~~~~~~~~~~~~~~~~~~~~~~~~~~~~~~~~$\left(m_1=0,\,m_2=1\right)$~~~~~~~~~~~~$\left(m_1=2, \, m_2=0\right)$
  \caption{An example of MPS degeneration formula for $M=2$. The first and the second quiver on the righthand side correspond to the partition $\left(m_1=0,\,m_2=1\right)$ and $\left(m_1=2, \, m_2=0\right)$  of $M=2$, respectively.}\label{fig:MPSexample}
\end{figure}

As a gentle warmup to the more detailed calculations that follow, we begin with the case $r=0$ and dimension vector  $(2,1).$  As discussed in \S \ref{sec:grass}, the moduli space is the Grassmannian $Gr(2,k)$ and we aim to reproduce the associated index \eqref{eqgrassy}.

There are two partitions $m_*^{(1)}=(m_1=0,\, m_2=1)$ and $m_*^{(2)}=(m_1=2,\, m_2=0)$ appearing in the degeneration formula \eqref{MPSy} (see Figure \ref{fig:MPSexample}). The quiver moduli space for the first partition is $\mathbb{CP}^{2k-1}$ and we have
\begin{align}
&\Omega(\mathcal{M}_{m_*^{(1)},1}^k,y) =y^{-2k+1}+ y^{-2k+3}+\cdots + y^{2k-3} +y^{2k-1}=
y^{-2k+1}[2k]_{y^2}~.
\end{align}

For the second partition $m_*^{(2)}$ in Figure \ref{fig:MPSexample}, the integrand for the residue formula is given by the general formula \eqref{Z1loop}:\footnote{Here we use a slightly simplified convention on the superscripts for the gauge fugacity $v$ and flavor fugacities $\xi$'s compared with \S \ref{sec:derivation}. To match the two conventions, we have $v_{2}^{(1)}=v,$ $\xi^{(I=1,\ell=1)}_i = \xi^{(1)},$ and $\xi^{(2,1)}_i=\xi^{(2)}$.}
\begin{align}
\left( {1\over \sin z}\right)^2 \prod_{i=1}^k 
{ \sin(u -\xi^{(1)}_i +z )\over \sin (u -\xi^{(1)}_i) }\,
{ \sin(u - v- \xi^{(2)}_i +z) \over \sin (u -v- \xi^{(2)}_i) }~,
\end{align}
where $u,v$ stand for the gauge fugacities for the rightmost and the leftmost node in the three-node quiver of Figure \ref{fig:MPSexample}.  The remaining node is decoupled. Meanwhile, $\xi^{(1)}$ and $\xi^{(2)}$ are the $su(k)$ flavor fugacities for the lower and upper group of arrows in Figure \ref{fig:MPSexample}, respectively. From the analysis in \S \ref{sec:derivation}, we see that the Jeffrey-Kirwan rule instructs us to pick up one factor of $\sin(u-\xi_i^{(1)})$ and one factor of $\sin(u-v-\xi_i^{(2)})$, and then sum over the contributions from all such choices. 

\begin{figure}[h]
\begin{center}
\includegraphics[width=.3\textwidth]{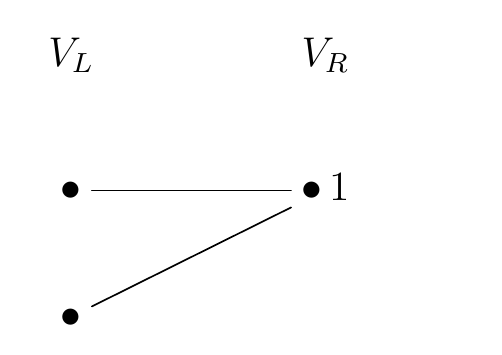}
\end{center}
\caption{The only tree in $\mathcal{G}(\vec{L}_{m_*^{(2)},0})$, where the vector $\vec{L}_{m_*^{(2)},0} = (0,0)$. In this example, $M=2$, $r=0$, and $m_*=\left( m_1=2,\,m_2=0\right)$. The number of vertices in $V_L$ and $V_R$ are $v_L=2$ and $v_R=1$. The number of edges is $e=2$.   Each vertex in $V_L$ is incident to $1$ edge. There is a unique tree in $\mathcal{G}(\vec{L}_{m_*^{(2)},0})$ corresponding to the poles $u-\xi^{(1)}_i=0,~u-v-\xi_j^{(2)}=0$ for different $\xi$'s.}\label{fig:j=0tree}
\end{figure}

In the language of bipartite graphs, this set of poles corresponds to a single tree in $\mathcal{G}(\vec{L}_{m_*^{(2)},0})$. Recall that a bipartite graph $G\in  \mathcal{G}(\vec{L}_{m_*^{(2)},0})$ has\footnote{ $M=2$ and $r=0$ here.}
\begin{align}
&v_L=\sum_\ell m_\ell=2~,~~~v_R =Mr+1= 1~,~~~e=Mr+\sum_\ell m_\ell =2~,
\end{align}
where $v_L$ and $v_R$ are the number of vertices in $V_L$ and $V_R$ and $e$ is the total number of edges in $G$.  The 2-dimensional vector is $\vec{L}_{m_*^{(2)},0}= (0,0)$. Each vertex in $V_L$ is incident to $1$ edge. The number of trees in $\mathcal{G}(\vec{L}_{m_*^{(2)},0})$ is given by the Tree Counting Theorem \eqref{Tmj} 
\begin{align}
T(\vec{L}_{m_*^{(2)} ,0} ) =1~.
\end{align}
This unique tree is illustrated in Figure \ref{fig:j=0tree}.

We now sum over these poles with different choices of $\xi$.  The resulting expression is compactly stated in terms of variables $\xi^{(I)}_{mi} =\xi^{(I)}_m-\xi^{(I)}_i$ and the function 
\begin{align}\label{fdef}
f(x)\equiv {\sin (x+z)\over \sin(x)}~.
\end{align}
We find
\begin{align}\label{flavorstep}
\Omega(\mathcal{M}^k_{m_*^{(2)},1},y) = \sum_{m,n=1}^k \prod_{i=1,\, i\neq m}^k 
f(\xi^{(1)}_{mi})\,
\prod_{\ell=1,\, \ell\neq n}^k 
f(\xi^{(2)}_{n\ell})~.
\end{align}

Despite its appearance,  the index \eqref{flavorstep} is in fact independent of the flavor fugacities $\xi$ due to $\mathcal{N}=4$ supersymmetry \cite{Hori:2014tda}. We can therefore evaluate $\Omega(\mathcal{M}^k_{m_*^{(2)},1},y)$ in a particular limit of $\xi$'s and obtain the exact answer as a function of $y$ (or equivalently, $z$). We choose the $\xi$'s to be purely imaginary 
\begin{align}
&\xi^{(I)} _{m} = -i \mathbb{R}_+,~~~~I=1,2~,
\end{align}
and further order them so that
\begin{align}\label{limit0}
\begin{split}
&i\xi^{(I)}_1 \gg i\xi^{(I)}_2\gg\cdots \gg i\xi^{(I)}_k\gg 1~,~~~~I=1,2~,\\
&i\xi^{(1)}_{mm'}\gg i\xi^{(2)}_{nn'}~,~~~~~~~~~~~~~~~~~~~~~~~\forall \,m<m'~,\,n<n'~.
\end{split}
\end{align}
In this limit we can replace the function $f(x)$ by the simpler expression
\begin{align}\label{limit}
f(\xi^{(I)}_{mm'})= {\sin(\xi^{(I)}_{mm'}+z)\over \sin(\xi^{(I)}_{mm'})} \rightarrow 
y^{\theta(m-m')}=
\begin{cases}
 y~,~~~~~~~~\text{if}~~m<m'~,\\
y^{-1}~,~~~~~\text{if}~~m>m'~,
\end{cases}
\end{align}
where $\theta(x)$ a step function 
\begin{align}\label{stepdef}
\theta(x) = 
\begin{cases}
 +1~,~~~~~~\text{if}~~x>0~,\\
-1~,~~~~~~\text{if}~~x<0~.
\end{cases}
\end{align}

In this limit we therefore have
\begin{align}
\sum_{m=1}^k \prod_{i=1,\, i\neq m}^k 
f(\xi_{mi}^{(I)}) 
=y^{-(k-1)} + y^{-(k-3)} +\cdots + y^{k-1}
= y^{-(k-1)} [k]_{y^2}~.
\end{align}
In fact, one can check that this trigonometric identity is true for arbitrary $\xi$'s. It follows that the refined index for the star quiver associated to the second partition $m_*^{(2)}$ is
\begin{align}
\Omega(\mathcal{M}^k_{m_*^{(2)},1},y)  =(y^{-(k-1)} [k]_{y^2})^2~.
\end{align}

Assembling the pieces using the degeneration formula \eqref{MPSy}, we find that the refined index for the Kronecker quiver with dimension vector $(2,1)$ and $k$ arrows is 
\begin{align}\label{Gr(2,k)}
\begin{split}
\Omega(2,1,k,y)&= y^{-2k+2}\left(  -{1\over 2[2]_{y^2}} y^{2k-1} \Omega(\mathcal{M}_{m_*^{(1)},1}^k,y)+{1\over 2} y^{2k-2}\Omega(\mathcal{M}_{m_*^{(2)},1}^k,y)  \right)\\
&={y^{2(2-k)}\prod_{i=1}^k (1-y^{2i})\over \prod_{i=1}^2 (1-y^{2i})\prod_{i=1}^{k-2} (1-y^{2i})}~.
\end{split}
\end{align}
This is indeed the expected answer \eqref{eqgrassy} for the Grassmannian $Gr(2,k)$.\footnote{If $k<2$ the index vanishes. }

\subsection{Example: $(2,2r+1)$}

We now generalize the calculations of the previous example to compute the refined index for the Kronecker quiver with dimension vector $(2,2r+1)$ and $k$ arrows. 

The index for the first partition $m_*^{(1)}= \left(m_1=0,\, m_2=1\right)$ in Figure \ref{fig:MPSexample} is that for the Grassmannian $Gr(2r+1,2k)$,
\begin{align}\label{Gr2r+1}
\Omega(\mathcal{M}_{m_*^{(1)},2r+1}^k,y) = y^{-(2r+1)(2k-2r-1)}
{\prod_{i=1}^{2k} (1-y^{2i}) \over\prod_{i=1}^{2r+1} (1-y^{2i}) \prod_{i=1}^{2k-(2r+1)} (1-y^{2i}) }~,
\end{align}
if $2k\ge 2r+1$ and zero otherwise. Note the exponent $(2r+1)(2k-2r-1)$ is the complex dimension of the $Gr(2r+1,2k)$.

\begin{figure}[h]
\begin{center}
\includegraphics[width=.4\textwidth]{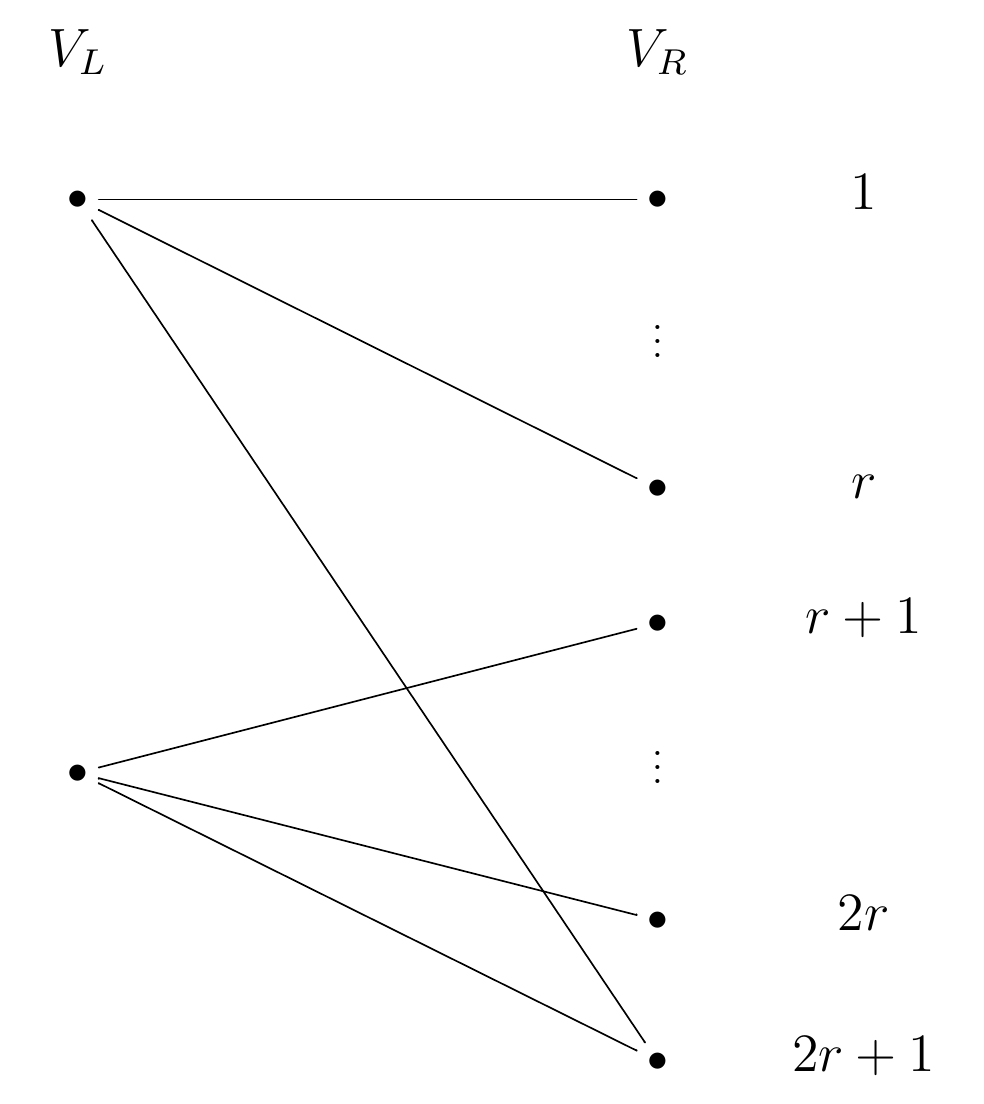}
\end{center}
\caption{A tree in $\mathcal{G}(\vec{L}_{m_*^{(2)},r})$, where $\vec{L}_{m_*^{(2)},r} = (r,r)$. In this example, $M=2$ and $m_*=\left( m_1=2,\,m_2=0\right)$. The first vertex in $V_L$ is incident to  vertices from 1 to $r$ and $2r+1$ in $V_R$, while the second vertex in $V_L$ is incident to  vertices from $r+1$ to $2r+1$ in $V_R$. There are $ (2r+1) { 2r\choose r}$ such trees, each of which corresponds to a set of poles like \eqref{pole}.}\label{fig:2r+1tree}
\end{figure}

For the second partition $m_*^{(2)}=\left(m_1=2,\,m_2=0\right)$, the type of bipartite graphs $G=(V_L+V_R,E)\in\mathcal{G}(\vec{L}_{m_*^{(2)},r})$ defined in \S \ref{subsec:graph} has\footnote{ $M=2$ here.}
\begin{align}
&v_L=\sum_\ell m_\ell=2~,~~~v_R = Mr+1= 2r+1~,~~~e=Mr+\sum_\ell m_\ell =2r+2~,
\end{align}
where $v_L$ and $v_R$ are the number of vertices in $V_L$ and $V_R$ and $e$ is the total number of edges in $G$. The 2-dimensional vector is $\vec{L}_{m_*^{(2)},r}= (r,r)$. Each vertex in $V_L$ is incident to $r+1$ edges. The number of trees in $\mathcal{G}(\vec{L}_{m_*^{(2)},r})$ is given by the Tree Counting Theorem \eqref{Tmj}
\begin{align}
T(\vec{L}_{m_*^{(2)} ,r} ) = (2r+1) { 2r\choose r}~.
\end{align}
Each tree corresponds to a set of poles with different flavor fugacities $\xi$ that contributes to the index in the residue formula.  Due to the Weyl symmetry permuting the $N$ vertices in $V_R$, each trees contributes the same amount to the index, so it suffices to compute one specific tree and multiply it by $T(\vec{L}_{m_*^{(2)},r})$. For definiteness, let us consider the tree in Figure \ref{fig:2r+1tree}, where the first vertex in $V_L$ is incident to vertices from 1 to $r$ and $2r+1$ in $V_R$, while the second vertex in $V_L$ is incident to  vertices from $r+1$ to $2r+1$ in $V_R$.

The tree in Figure \ref{fig:2r+1tree} corresponds to the set of poles specified by
\begin{align}\label{pole}
\begin{split}
&u_{2r+1}=\xi^{(1)}_{\mu_0}~,~~
~~\,~~~\hspace{1pt}~u_1 = \xi^{(1)}_{\mu_1}~ ,~~
~~~\,~~~\hspace{.5pt}~~~u_2= \xi^{(1)}_{\mu_2}~,~~\cdots,~~
~~~~~\,~\hspace{2pt}~u_r =\xi^{(1)}_{\mu_r}~,\\
&u_{2r+1}-v=\xi^{(2)}_{\nu_0}~,~~~\,
u_{r+1}-v=\xi^{(2)}_{\nu_1}~,~~~\,
u_{r+2}-v=\xi^{(2)}_{\nu_2}~,~~
\cdots,~~
u_{2r} -v= \xi^{(2)}_{\nu_r}~,
\end{split}
\end{align}
for some  $\mu_0,\,\mu_1,\cdots,\,\mu_r,\,\nu_0,\, \nu_1,\cdots,\,\nu_r=1,\cdots,k$. The $\mu$'s are all distinct and similarly for $\nu$'s. In the end we sum over all such $\mu$'s and $\nu$'s. Note that the above pole prescription fixes 
\begin{align}
v= \xi^{(1)}_{\mu_0} - \xi^{(2)}_{\nu_0}~.
\end{align}

The integrand for the Jeffrey-Kirwan residue is given by \eqref{Z1loop}
\begin{align}
\left( {1\over \sin(z)}\right)^{2r+2}
\prod_{\substack{a,b=1\\a\neq b}}^{2r+1} {1\over f(u_a-u_b)}
\prod_{i=1}^k\prod_{a=1}^{2r+1} f(u_a-\xi^{(1)}_i)
f(u_a-v-\xi^{(2)}_i)~,
\end{align}
where  $f(x) $ is defined by \eqref{fdef}. The arguments of $f(x)$ in the integrand are determined by the position of the pole \eqref{pole}. They are
\begin{align}
\begin{split}
&u_c-\xi^{(1)}_i = \xi^{(1)}_{\mu_ci}~,~~~~~~~~~~~~~~~~~~~~~~~
u_{r+c}-\xi^{(1)}_i = \xi^{(1)}_{\mu_0 i } +\xi^{(2)}_{\nu_c \nu_0}~,\\
&u_c-v-\xi^{(2)}_i = \xi^{(1)}_{\mu_c\mu_0}+\xi^{(2)}_{\nu_0 i}~,
~~~~~~~~u_{r+c}-v-\xi^{(2)}_i = \xi^{(2)}_{\nu_c i}~,\\
&u_{2r+1}- \xi^{(1)}_i = \xi^{(1)}_{\mu_0i}~,
~~~~~~~~~~~~~~~~~~~u_{2r+1}-\xi^{(2)}_i = \xi^{(2)}_{\nu_0i}~,
\end{split}
\end{align}
where $c,d=1,2,\cdots,r$ (rather than to $2r+1$).

Since the refined index does not depend on the flavor fugacities $\xi$'s by the $\mathcal{N}=4$ supersymmetry, we are again free to evaluate it in the limit \eqref{limit0}. In this limit $f(x)f(-x)=1$, so we do not have to consider the factor $\prod_{a\neq b}^{2r+1}{1\over f(u_a-u_b)}$ in the integrand. We can also replace $f(\xi^{(1)}_{\mu_c\mu_0} +\xi^{(2)}_{\nu_0 i})$ by $f(\xi^{(1)}_{\mu_c\mu_0})$.  The index is then
\begin{align}
\begin{split}
\Omega(\mathcal{M}_{m_*^{(2)},2r+1}^k,y)
 =& {1\over r!r!} \sum_{(\mu_0,\,\mu_1,\,\cdots,\,\mu_r)}
\left(\prod_{\substack{i=1\\i\neq \mu_0}}^k
f(\xi^{(1)}_{\mu_0i})^{r+1}
\right)
\left(
\prod_{c=1}^r 
\prod_{\substack{i=1\\i\neq \mu_c}}^k
f(\xi^{(1)}_{\mu_ci})
\right)
\left( \prod_{c=1}^r f(\xi^{(1)}_{\mu_c\mu_0})^k\right)\\
&\times
\sum_{(\nu_0,\,\nu_1,\,\cdots,\,\nu_r)}
\left(
\prod_{\substack{i=1\\i\neq \nu_0}}^k
f(\xi^{(2)}_{\nu_0i})
\right)
\left(
\prod_{c=1}^r 
\prod_{\substack{i=1\\i\neq \nu_c}}^k
f(\xi^{(2)}_{\nu_ci})
\right)
\left(\prod_{c=1}^r f(\xi^{(2)}_{\nu_c\nu_0})\right)~,
\end{split}
\end{align}
where the sum is over all tuples $(\mu_0,\,\mu_1,\,\cdots,\,\mu_r)$ with distinct $\mu$'s ranging from $1$ to $k$. Similarly for $(\nu_0,\,\nu_1,\,\cdots,\,\nu_r)$.

Using \eqref{limit}, we have
\begin{align}
\begin{split}
&\prod_{\substack{i=1\\i\neq \mu_a}}^k
f(\xi^{(1)}_{\mu_a i}) 
=y^{k-2\mu_a+1}~.
\end{split}
\end{align}
We can then rewrite the index for the star quiver associated to $m_*^{(2)}$ as
\begin{align}
\begin{split}
\Omega(\mathcal{M}_{m_*^{(2)},2r+1}^k,y)=&
{1\over r!r!}
\left(
\sum_{(\mu_0,\,\mu_1,\,\cdots,\,\mu_r)} 
y^{ (2r+1)(k+1) -2(r+1)\mu_0 - 2 \sum_{c=1}^r \mu_c - k\sum_{c=1}^r\theta(\mu_c-\mu_0)  } 
\right)\\
&\times
\left(
\sum_{(\nu_0,\,\nu_1,\,\cdots,\,\nu_r)}
y^{ (r+1)(k+1) -2 \nu_0 -2 \sum_{c=1}^r \nu_c -\sum_{c=1}^r \theta(\nu_c-\nu_0) }
\right)~.
\end{split}
\end{align}
Again the sum is over all tuples $(\mu_0,\,\mu_1,\,\cdots,\,\mu_r)$ with distinct $\mu$'s ranging from $1$ to $k$. Similarly for $(\nu_0,\,\nu_1,\,\cdots,\,\nu_r)$, and $\theta$ is the step function defined in \eqref{stepdef}.

Assembling the pieces using the degeneration formula \eqref{MPSy}, we arrive at our final expression for the refined index of the Kronecker quiver with dimension vector $(2,2r+1)$
\begin{align}
\begin{split}
\Omega(2,2r+1,k,y)&=
y^{-2-d}\Big[
-{1\over 2(1+y^2)} 
y^{d_{m_*^{(1)}} } \Omega(\mathcal{M}^k_{m_*^{(1)},2r+1} , y)
+{1\over 2}y^{d_{m_*^{(2)}}} \Omega(\mathcal{M}_{m_*^{(2)},2r+1}^k,y)
\Big]~,
\end{split}
\end{align}
if $d=(2r+1)(2k-2r-1)-3\ge 0$ and zero otherwise. Here $d_{m_*^{(1)}} = (2r+1)(2k-2r-1)$ and $d_{m_*^{(2)}} = 2k(2r+1) -(2r+1)^2-1$ are the ranks of the star quivers associated to the partitions $m_*^{(1)}$ and $m_*^{(2)}$, respectively. $\Omega(\mathcal{M}^k_{m_*^{(1)},2r+1},y)$ is the refined index for the Grassmannian $Gr(2j+1,2k)$ given in \eqref{Gr2r+1}. 

We list the first few examples of the refined index $\Omega(M,N,k,y)$ in the following:\footnote{For each $k,$ the remaining cases not listed in \eqref{spinanswers} are redundant as a consequence of the isomorphism \eqref{iso1} $\Omega (2,\, 2r+1,\,k) = \Omega ( 2, \, 2(k-r-1)+1,\,k ).$}
\begin{align*}\label{spinanswers}
\begin{split}
&k=2:\\
&~~\Omega(2,1,2 ,y) = 1~,\\
&k=3:\\
&~~\Omega(2,1,3 ,y) =  {1\over y^2} +1 +y^2~,\\
&~~\Omega(2,3,3,y) = {1\over y^6} + {1\over y^4} + {3\over y^2} +3 +3y^2 +y^4 +y^6~,\\
&k=4:\\
&~~\Omega(2,1,4 ,y) = {1\over y^4}  + {1\over y^2}  + 2 +y^2 + y^4~,\\
&~~\Omega(2,3,4 ,y) = {1\over y^{12}} + {1\over y^{10}} +{3\over y^8} + {4\over y^6} + {7\over y^4} +{8\over y^2} + 10+8y^2 + 7y^4 + 4y^6 + 3y^8 + y^{10} +y^{12}~,\\
&k=5:\\
&~~\Omega(2,1,5 ,y)= \frac{1}{y^6}+\frac{1}{y^4}+\frac{2}{y^2}+2+2 y^2+y^4+ y^6~,\\
&~~\Omega(2,3,5 ,y) = {1\over y^{18}} + {1\over y^{16}} +{3\over y^{14}} +{4\over y^{12}} +{7\over y^{10}} + {9\over y^8} + {14\over y^6}
  +{16\over y^4} +{20\over y^2}+20 \\
  &~~~~~~~~~~~~~~~~~~+ 20y^2  + 16y^4 + 14y^6 + 9y^8  + 7y^{10} +4y^{12} +3y^{14}+y^{16} +y^{18}~,\\
&~~\Omega(2,5,5 ,y) = \frac{1}{y^{22}}+\frac{1}{y^{20}}+\frac{3}{y^{18}}+\frac{4}{y^{16}}+\frac{8}{y^{14}}+\frac{11}{y^{12}}+\frac{17}{y^{10}}+\frac{22}{y^8}+\frac{30}{y^6}+\frac{35}{y^4}+\frac{41}{y^2}+41\\
   &~~~~~~~~~~~~~~~~~~+41 y^2+35 y^4+30 y^6+22 y^8+17 y^{10}+11 y^{12}+8 y^{14}+4 y^{16}+3 y^{18}+y^{20}+y^{22}~.\\
  \end{split}
\end{align*}
We have checked that the above refined indices agree with \cite{2003InMat.152..349R}.

\section*{Acknowledgements} 
We would like to thank Murad Alim, Melody Chan, Tiffany Yu-Han Chen, Yu-Wei Fan, Ben Heidenreich, Kazuo Hosomichi, Daniel Jafferis, Cumrun Vafa, and Fan Wei for many useful discussions. We are especially grateful to Yi-Hsiu Chen, Noam D. Elkies, Ira Gessel, Ying-Hsuan Lin, and Alexey Ustinov for many crucial steps in the proofs. SHS is grateful for the Kavli Institute for the Physics and Mathematics of the Universe (IPMU) and National Taiwan University's hospitality during the final stage of the work. The work of CC is support by a Junior Fellowship at the Harvard Society of Fellows.  The work of SHS is supported by the Kao Fellowship at Harvard University.

\appendix

\section{An Identity for $[x^{M}]\{\exp(\beta F)\}$ When $r=1$}
\label{sec:powerseries}

In this appendix, we provide a direct calculation of the power series coefficients of $ \exp \left[ \,\beta F(k,1,x) \,\right]$ for arbitrary complex $\beta.$

We first prove a lemma.\footnote{We thank Noam D. Elkies and Ira Gessel for pointing out this identity to us.}
\begin{lemma}\label{app:lemma}
Let $k$ and $M$ be positive integers and $F(k,1,x) = \sum_{\ell=1}^\infty {(-1)^{\ell-1}\over \ell^2} (\ell+1) { k \ell\choose \ell +1}x^\ell$. Then
\begin{align}
F(k,1,\, y(1+y)^{k-1} ) = k(k-1) \log (1+y)~.
\end{align}
\end{lemma}

\paragraph{Proof}

We will prove this by direct substitution. The lefthand side equals to
\begin{align}
F(k,1,\, y(1+y)^{k-1}) = \sum_{\ell=1}^\infty { (-1)^{\ell-1} \over\ell^2} (\ell+1) { k\ell \choose \ell+1} y^\ell\times \left[\,
 \sum_{s=0}^{\ell(k-1)}  { \ell (k-1)\choose s} y^s
 \,\right]~.
\end{align}
It follows that 
\begin{align}\label{A3}
\begin{split}
[y^n] \Big\{ \, F(k,1,\,y(1+y)^{k-1} ) \,\Big\}  &= \sum_{\ell=1}^n {(-1)^{\ell-1}\over \ell^2}(\ell+1)
{k\ell \choose \ell+1}  {  (k-1) \ell\choose n-\ell}\\
&=-{k(k-1)\over n!}  \, \sum_{\ell=1}^n (-1)^\ell { n\choose \ell} \prod_{i=1}^{n-1} (k\ell -i)~.
\end{split}
\end{align}
Note that for $m\in \mathbb{N}$,
\begin{align}
\left(x{d\over dx} \right)^m (1-x)^n = \sum_{\ell=1}^n (-1)^\ell { n\choose \ell}\, \ell^m\, x^n~.
\end{align}
Setting $x=1$, it follows that
\begin{align}\label{binomial}
\sum_{\ell=1}^n (-1)^\ell {n\choose \ell} \,\ell^m  = 0~~~~\text{if}~~n>m\ge1~.
\end{align}
Using the above identity, we see that all the higher order terms in $\ell$ in \eqref{A3} vanish after summing over $\ell$, and we are left with
\begin{align}
\begin{split}
[y^n] \Big\{ \, F(k,1,\,y(1+y)^{k-1} ) \,\Big\}  &=
- {k (k-1)\over n!} \sum_{\ell=1}^n (-1)^\ell {n\choose \ell} \prod_{i=1}^{n-1}(- i)\\
&=(-1)^{n-1} {k(k-1)\over n}~.
\end{split}
\end{align}
This concludes the proof of the lemma. $\square$

\begin{theorem}
Let $k$ and $M$ be positive integers and $F(k,1,x) = \sum_{\ell=1}^\infty {(-1)^{\ell-1}\over \ell^2} (\ell+1) { k \ell\choose \ell +1}x^\ell$. Let $\beta$ be a complex number. Then
\begin{align}\label{generalidentity}
[x^M] \Big\{ \exp \left[ \,\beta F(k,1,x) \,\right] \Big\}  =  {\beta\over M} k(k-1){ k(k-1)\beta-(k-1)M-1\choose M-1}~.
\end{align}
\end{theorem}

\paragraph{Proof}

From Lemma \ref{app:lemma}, we have
\begin{align}\label{expNF}
\exp \Big[ \, \beta F(k,1,x ) \,\Big] = \left(1+y(x)\right)^{\beta k(k-1)} = \sum_{\ell=0}^{\infty}{\beta k(k-1)\choose \ell} y(x)^\ell~,
\end{align}
where $y(x)$ is defined as the solution to
\begin{align}
x=y(1+y)^{k-1}.
\end{align}
The Taylor series coefficient of $y(x)^\ell$ can be obtained by the Lagrange inversion theorem (see, for example, \cite{Noam1}\cite{Noam2}),
\begin{align}
y(x)^\ell =  \sum_{m=0}^\infty(-1)^{m} { \ell\over m+\ell} { (k-1) (m+\ell) + m-1\choose m}x^{m+\ell}~.
\end{align}
Plugging this into \eqref{expNF}, we have
\begin{align}\label{A11}
[x^M]\Big\{ \exp \left[\, \beta F(k,1,x) \,\right] \Big\}&={1\over M}\sum_{\ell=1}^{M} (-1)^{M-\ell} {\ell}\,
{\beta k(k-1)\choose \ell}{kM-\ell-1\choose M-\ell}~.
\end{align}
Note that  right-hand-sides of both \eqref{generalidentity} and \eqref{A11} are polynomials in $\beta $ of degree $M$. To show that they are the same, it suffices to show that they have the same roots in $\beta $, while the overall constant can be trivially checked by plugging some specific values of $M,\beta ,k$. Due to the binomial coefficient, \eqref{generalidentity} has zeroes at $\beta =0$ and
\begin{align}\label{zeroes}
\beta ={1\over k(k-1)} (kM -s )~ ,~~~~s=1,\cdots,M-1~.
\end{align}
Obviously \eqref{A11} vanishes when $\beta =0$. It remains to show that \eqref{A11} vanishes when $\beta $ takes value in \eqref{zeroes}. When $k(k-1)\beta  =  (k-1)M - s$, \eqref{A11} can be written as
\begin{align}\label{A14}
&{(-1)^M\over M\times M!} \sum_{\ell=1}^M (-1)^\ell \ell { M \choose \ell}  { \left( kM-s\right)!\over \left( kM-(\ell+s)\right)! } 
{\left( kM-(\ell+1)\right)!\over  \left( kM - M-1\right)!}\\
&={(-1)^M\over M\times M!} {\left( kM-s\right)!\over  \left( kM - M-1\right)!}
\sum_{\ell=1}^M (-1)^\ell {M\choose \ell} \, \ell\left(kM-(\ell +1)\right) \cdots
\left(kM-(\ell +s-1)\right)~.
\end{align}
Note $\ell\left(kM-(\ell +1)\right) \cdots\left(kM-(\ell +s)\right)$ is a polynomial in $\ell$ of degree $s$. Since $s< M$, it follows from \eqref{binomial} that \eqref{A14} is zero. Thus we have shown that $\beta ={1\over k(k-1)} (kM-s)$ with $s=1,\cdots, M-1$ and $\beta =0$ are roots for \eqref{A11}, which are the same $M$ roots of \eqref{generalidentity}. This completes the proof. $\square$

\section{Degenerate Poles}\label{sec:degenerate}

In \S \ref{sec:derivation} we have computed the contributions to the index from the non-degenerate poles in the residue formula \eqref{JKform}. Here, we argue that the degenerate poles do not contribute when $N=Mr+1$.

Consider an integrand of the following form
\begin{align}\label{degint}
{\sin\left[ A^{(1)}(u)\right] \cdots  \sin\left[ A^{(q)}( u)\right]\over \sin\left[B^{(1)}(u)\right]\cdots \sin\left[B^{(p)}(u)\right] } \cdots~,
\end{align}
where $A^{(i)}(u)$ and $B^{(i)}(u)$ are some linear functions of the gauge fugacities\footnote{As an abuse of notations, in the general discussion here we use $u_\alpha$ to represent both the gauge fugacities of the non-abelian node $u_a$ and of the abelian nodes $v_I^{(\ell)}$ in the star quiver (Figure \ref{fig:star}).} $u_\alpha$. The $\cdots$ represents terms that are nonzero and finite at the pole. The index $\alpha$ runs over $1,\cdots, e$, where $e$ is the total rank of the gauge group. Suppose  at a point $u=u_*$, we have
\begin{align}
A^{(i)}( u_{*}) = 0~,~~~B^{(i)}( u_{*})=0~.
\end{align}
Then $u=u_*$ is called a degenerate pole if $p>e$, and a non-degenerate  pole if $p=e$ and $q=0$. For the degenerate poles in the star quiver, we always have $p-q=e$, so they are effectively simple poles after carefully ``canceling" the zeroes in the numerator and the denominator. We will be precise about this cancellation in a moment.

Let $u_*$ be a degenerate pole with $p-q=e$. As far as the residue at $u=u_*$ is concerned, we can replace  $\sin\left[A^{(i)}(u)\right]$ and $\sin\left[B^{(i)}(u)\right]$ by their arguments.\footnote{ It is not, however, legitimate to replace the sine factors in $\cdots$ of \eqref{degint} by their arguments as they do not vanish at $u_*$.} After doing so, we  perform partial fraction decomposition to break \eqref{degint} into simple poles:
\begin{align}\label{PFD}
{ A^{(1)}( u) \cdots  A^{(q)}( u) \over B^{(1)}( u)\cdots B^{(p)}( u)} = \sum_w {1\over C^{(1,w)}( u)\cdots C^{(e,w)}(u)}~,
\end{align}
for some linear functions $C^{(a,w)}(u)$ of $u_\alpha$.  Now the terms on the righthand side become non-degenerate poles and we can evaluate their residue at $u=u_*$ using the technique in \S \ref{sec:JKrule}. In other words, \textit{we have decomposed a degenerate pole with $p-q=e$ into a sum of non-degenerate poles.}

In the star quiver $\mathcal{M}^k_{m_*, N}$ with $N=Mr+1$, the degenerate poles decompose into non-degenerate poles as described above. We claim that none of these non-degenerate poles satisfies the Jeffrey-Kirwan rule of \S\ref{sec:JKrule} and therefore they do not contribute to the index.

Let us demonstrate this claim in an explicit example. Let $M=3$, $r=1$, $e=6$, and $m_* = (m_1=3, \,m_2=0,\,m_3=0)$. An example of a degenerate pole is\footnote{For simplicity of notation, we will denote $v_I^{\ell=1}$ simply by $v_I$ as there are no other values of $\ell$ for the partition $m_*= (m_1=3, \,m_2=0,\,m_3=0)$. We also abbreviate $\xi_i^{(I,\ell=1)}$ by $\xi_i^{(I)}$. As usual $v_1= v_1^{(1)}$ is understood to be zero since it corresponds to the decoupled abelian node.}
\begin{align}\label{degpole}
\begin{split}
& u_1 -v_1 -\xi^{(1)}_1 =0~,~~~~~
u_2-v_1-\xi^{(1)}_1=0~,\\
&u_1-v_2-\xi^{(2)}_1=0~,~~~~~
u_2-v_2-\xi^{(2)}_1=0~,~~~~~
u_3-v_2-\xi^{(2)}_2=0~,\\
&u_1-v_3-\xi^{(3)}_1=0~,~~~~~ 
u_2-v_3-\xi^{(3)}_1=0~,~~~~~
u_4-v_3-\xi^{(3)}_2=0~.
\end{split}
\end{align}
These correspond to the $p=8$ sine factors that vanish at $u=u_*$ in the denominator of \eqref{degint}. Note that since $u_1-v_2-\xi^{(2)}_1=0$ and $u_2-v_2-\xi^{(2)}_1=0$, we have $u_1=u_2$ at this pole. This results in $q=2$ sine factors that vanish at $u=u_*$ in the numerator of \eqref{degint},
\begin{align}
\sin(u_1-u_2)\sin(u_2-u_1)~.
\end{align}
Note that indeed we have $p-q=e$. 

Now performing the partial fraction decomposition, there are four terms on the righthand side of \eqref{PFD}, each of which corresponds to a non-degenerate pole. For example, one of them is
\begin{align}
\begin{split}
&u_1-v_1-\xi^{(1)}_1=0~,\\
&u_1-v_2-\xi^{(2)}_1=0~,~~~~~u_3-v_2-\xi^{(2)}_2=0~,\\
&u_1-v_3-\xi^{(3)}_1=0~,~~~~~u_2-v_3-\xi^{(3)}_1=0~,~~~~~u_4-v_3-\xi^{(3)}_2=0~.
\end{split}
\end{align}
However, this non-degenerate pole does not satisfy the Jeffrey-Kirwan rule. For example, there should be $\ell r+1=2$ sine factors with arguments $u_a-v_1-\xi$, whereas there is only one above. It follows that this non-degenerate pole does not contribute to the index. Similarly one can show the other three non-degenerate poles also violate the Jeffrey-Kirwan rule.  Hence the degenerate pole \eqref{degpole}, which equals to the sum of four non-degenerate poles, does not contribute to the index.

This argument may be generalized to demonstrate that in general, degenerate poles do not contribute to the index when $N=Mr+1.$

\bibliographystyle{utphys}
\providecommand{\href}[2]{#2}\begingroup\raggedright
\end{document}